\documentclass[fleqn,usenatbib]{mnras}
\usepackage{lmodern}
\usepackage{float}
\usepackage[all]{hypcap}
\usepackage{color,soul}

\usepackage[normalem]{ulem}

\usepackage{newtxtext,newtxmath}
\usepackage{anyfontsize}
\usepackage[T1]{fontenc}
\usepackage{caption}
\usepackage[normalem]{ulem}
\usepackage{amsmath,siunitx}

\defcitealias{Gurkan2018}{G18}	
\defcitealias{2023MartinH_optcounterparts}{H23}	
\defcitealias{Smith20}{S21}	
\defcitealias{drake2024}{Dr24}	
\defcitealias{das2024}{D24}	

\DeclareRobustCommand{\VAN}[3]{#2}
\let\VANthebibliography\thebibliography
\def\thebibliography{\DeclareRobustCommand{\VAN}[3]{##3}\VANthebibliography}

\usepackage{graphicx}	
\usepackage{amsmath}

\title[What determines the RC--SFR in LoTSS-DR2?]
{What factors shape the radio luminosity of star-forming galaxies? A new calibration from LoTSS-DR2}

\author[S. Shenoy et al.]{Shravya Shenoy$^{1}$,\thanks{E-mail: shravyashenoy.astro@gmail.com }
Daniel J. B. Smith$^{1}$,
Sarah K. Biddle$^{2}$,
G\"ulay G\"urkan$^{1,3}$,
Martin J. Hardcastle$^{1}$,
\newauthor
Marina I. Arnaudova$^{1,4}$,
Soumyadeep Das$^{1}$,
Luke R. Holden$^{1}$,
Gaoxiang Jin$^{5}$,
Leah K. Morabito$^{6,7}$,
\newauthor
Huub J. A. R\"ottgering$^{8}$
\\
$^{1}$Centre for Astrophysics Research, University of Hertfordshire, Hatfield, AL10 9AB, UK\\
$^{2}$Harvard-Smithsonian Center for Astrophysics, 60 Garden St., Cambridge MA 02138, USA\\
$^{3}$Australia Telescope National Facility, CSIRO Space and Astronomy, PO Box 1130, Bentley WA 6102, Australia\\
$^{4}$Institute for Astronomy, University of Edinburgh, Royal Observatory, Blackford Hill, Edinburgh, EH9 3HJ, UK.\\
$^{5}$Max Planck Institute for Astrophysics, Karl-Schwarzschild-Str. 1, D-85741 Garching, Germany\\
$^{6}$Centre for Extragalactic Astronomy, Department of Physics, Durham University, South Road, Durham DH1 3LE, UK \\
$^{7}$Institute for Computational Cosmology, Department of Physics, Durham University, South Road, Durham DH1 3LE, UK\\
$^{8}$Leiden Observatory, Leiden University, PO Box 9513, 2300 RA Leiden, The Netherlands\\
}

\date{Accepted XXX. Received YYY; in original form ZZZ}

\pubyear{2025}

\begin{document}
\label{firstpage}
\pagerange{\pageref{firstpage}--\pageref{lastpage}}
\maketitle

\begin{abstract}
Radio observations offer a dust-unobscured view of galaxy star formation via the radio continuum-star formation rate (RC--SFR) relation. Emerging evidence of a stellar mass dependence in the RC--SFR relation raises the broader question of how other galaxy properties may influence this relation. In this work, we study the dependence of the global RC--SFR relation on galaxy properties in local ($z\,\leq$\,0.3) star-forming galaxies (SFGs) using the second data release of the LOFAR Two-Metre Sky Survey (LoTSS-DR2). Employing a non-parametric decision-tree regression algorithm, we identify the most important galaxy properties for estimating the radio luminosity using a sample of 18,828 emission-line-classified SFGs based on spectroscopic data from the SDSS-DR8. Along with the spectroscopically obtained SFRs and stellar mass values, we also use SFRs and stellar masses derived using photometric SED-fitting from the \textit{GALEX}--SDSS--\textit{WISE} Legacy Catalogue (GSWLC) for the same sample. We find that a galaxy's SFR is most important for predicting the radio luminosity, followed by the stellar mass, at $>5\sigma$ significance. Complementing the LoTSS catalogue 150\,MHz flux densities with aperture photometry for the rest of the emission-line classified sample (35,099 galaxies in total), we obtain a new calibration of the RC--SFR relation, which does not change significantly whether we use spectroscopic or photometrically derived SFRs and stellar masses, despite the fact that the methods probe star formation on different characteristic timescales. Our results highlight the utility of decision-tree algorithms for handling censored radio-selected galaxy samples, which will be useful for future spectroscopic surveys of radio sources.

\end{abstract}


\section{Introduction}

Exploring star formation in galaxies is important for understanding the evolution of baryonic matter. Traditionally, star formation rate (SFR) calibrations relied on ultraviolet and optical observations of ionising radiation produced by young and massive stars, because they dominate the total stellar luminosities of star-forming galaxies (see \citealt{2012KennicuttEvans}; \citealt{Calzetti13} and \citealt{MadauDickinson14} for reviews). However, reliably using optical and UV data to determine the intrinsic SFRs of galaxies is challenging, especially in the distant universe, due to the large and uncertain dust corrections required (e.g. \citealt{1998AKennicutt_SF}). Dust absorbs a fraction of the UV\slash optical radiation arising from short-lived luminous stars, and the energy is then re-radiated at far-infrared (FIR) wavelengths, where the luminosity is therefore a generally reliable means of estimating SFRs in dusty galaxies (e.g. \citealt{1987Perssonlonsdale,1990DevereuxYoung};
\citealt{1998AKennicutt_SF,2012KennicuttEvans}). However, along with the practical difficulties associated with obtaining FIR observations in the post-\textit{Herschel} era, there are also issues arising due to the low spatial resolution of FIR data \citep[e.g.][]{smith2011,mccheyne2022}, as well as the possible impact of unobscured sight-lines \citep[e.g.][]{2006Perez,sorba2015,smith2018,haskell2023} and additional heating from evolved stellar populations \citep[e.g.][]{1987Perssonlonsdale}.
The dust that affects short-wavelength star formation indicators is irrelevant at radio frequencies, where a tight correlation between radio and far-infrared luminosities has been observed (e.g. \citealt{1971Vanderkruit,1985DeJong};
\citealt{1985Helou};
\citealt{2010Jarvis,2010Ivison,2014smithfirrc,2018SRead,2021Molnar}). Non-thermal radio emission in SFGs is attributed to electrons accelerated by supernova shocks moving in the ambient magnetic field \citep{2004Berezhko}. Massive stars $(\gtrsim  8\,{M}_{\odot}$) are responsible for core-collapse supernovae \citep{Condon1992}, and so the observed non-thermal radio emission can be used as an indicator of the recent ($\simeq 100$\,Myr averaged) star formation rate of galaxies.\\
\\
Multiple observational works have studied the relationship between galaxies' SFRs and their non-thermal synchrotron radio emission at 150\,MHz including \citet[hereafter \citetalias{Gurkan2018}]{Gurkan2018}; \citet{2019Wang}; \citet[hereafter \citetalias{Smith20}]{Smith20}; \citet{2021Bonato}; \citet{2022Heesenrcsfr}; \citet{2023Best} and \citet[hereafter \citetalias{das2024}]{das2024}. These works assumed a power law relation between SFR and radio luminosity (at various frequencies), with the gradient often deviating from linearity. A gradient equal to one would mean that the galaxy is a perfect electron calorimeter and that the cosmic ray electrons lose all their energy before being transported away from the sites at which radiation is observed. However, the radio emission may underestimate the SFRs of galaxies compared to far-infrared tracers when a fraction of cosmic rays accelerated by supernovae escape the host galaxy \citep{Condon1991}. At the same time, massive galaxies could retain cosmic ray electrons more efficiently due to deeper potentials, overestimating star-formation rates in galaxies and so giving rise to a super-linear slope of the RC--SFR power-law relation as observed e.g. by \citet{Davies2017}. \citet{2019Wang} report values as steep as $\sim$ 1.35 using an infrared (60\,$\mu$m) selected subsample of LOw-Frequency ARray Two-metre Sky Survey (LoTSS; \citealt{2017Shimwell}) galaxies at 150\,MHz. On spatially-resolved scales using integral field spectroscopy (IFS)-derived SFR, \citet{2024jin_rcsfr} report a superlinear slope of 1.16. \\
\\ 
Works including \citetalias{Gurkan2018} and \citetalias{Smith20} have explored the dependence of the RC--SFR relation on stellar mass, demonstrating that including a mass-dependence reduces the scatter on the RC--SFR relation. Possible interpretations for stellar mass dependence of the RC--SFR relation are that more massive galaxies are better cosmic ray calorimeters or that stellar mass dependence comes from undiagnosed contamination by active galactic nuclei (AGN), since it is well-known that AGN are also associated with non-thermal GHz emission \citep[e.g.][]{best2012} and that the AGN fraction is dependent on stellar mass \citep{2019Sabater}. However, the stellar mass dependence in \citetalias{Gurkan2018} and \citetalias{Smith20} extends to relatively low-mass galaxies ($<10^{10}\,{M}_{\odot}$) where the AGN fraction is small (\citealt{2019Wendy_optmatches}; \citealt{2021Mandal}; \citealt{2023Best}; \citealt{drake2024}; \citetalias{das2024}), consistent with results from simulations like The Tiered Radio Extragalactic Continuum Simulation or T-RECS \citep{2019bonaldi-trecs}. Therefore, AGN contamination is unlikely to be the only explanation for stellar mass dependence in the RC--SFR relation.\\
\\
The stellar mass dependence of the RC--SFR relation prompts a broader question: which additional galaxy properties might influence the relation? To address this question, we generate a non-parametric random forest (RF) regression model that predicts the radio luminosity of SFGs using galaxy properties (which can be derived using photometric, spectroscopic, or a mixture of the two types of observations). The application of RF regressors \citep{2001breiman} to explore fundamental scaling relations in extragalactic studies is becoming increasingly popular (see, e.g. \citealt{2019SanchezAlmeida};  \citealt{2020Bluck}; \citealt{Bluck2019}; \citealt{2022Piotrowska}; \citealt{2023bBaker}; \citealt{2023aBaker}; \citealt{2024Maheson}; \citealt{2024Sanchez} among others), since ensemble learning methods with decision trees like RF regression can identify highly non-linear relationships in multi-dimensional datasets. While RF regressors cannot predict analytical relations between the input and target variables directly, the relative importance of features that are included in the model can be estimated, leading to a better intuitive understanding of the model. \\ 
\\
In this work, we use a sample of 35,099 emission-line classified SFGs, of which 18,828 are detected at $\geq 3 \sigma$ at 150\,MHz. We apply RF regression to the radio-detected sources (18,828 sources) to investigate the dependence of radio luminosity on galaxy features including SFR, stellar mass, gas-phase metallicity and velocity dispersion. Using the results of the RF model, we then fit for an analytical relation to link the most important features to radio luminosity using the full sample of 35,099 galaxies, which includes radio non-detections. \\
\\
The paper is organised as follows: in Section \ref{data} we describe the data used in this work, while in Section \ref{selection} we outline our sample definition and present the properties of our sample. In Section \ref{results} we describe the methods used in this paper, and outline our results, including those derived using the RF method in Section \ref{ml} and our parametrization of the RC--SFR relation in Section \ref{rcsfrmass}. We discuss our results and present our conclusions in Section \ref{discussion}. Throughout this paper we adopt a flat $\Lambda$CDM cosmology with $H_0$=70\,km\,s$^{-1}$\,Mpc$^{-1}$, $\Omega_{m}$ = 0.3 and $\Omega_{\Lambda}$ = 0.7. Spectral index $\alpha$ is defined as S\,$\propto \nu^{\alpha}$ with a typical value of $\alpha$ = $-0.7$. `\texttt{log}' in this paper refers to \texttt{log}$_{10}$ with the exception of ln $\mathcal{L}$ which we call log-likelihood following standard terminology. We note that $\psi \equiv \mathrm{SFR}$ throughout this paper and we use O/H interchangeably with $12+\log(\mathrm{O/H})$ to denote the gas-phase oxygen abundance. All photometric magnitudes are in the AB system \citep{1983OkeGunn} unless otherwise specified. 

\section{Data}\label{data}
This work makes use of three primary datasets, as described in the following sections.
\subsection{LoTSS Data release 2}
\begin{figure}
    \centering
    \includegraphics[scale=0.5]{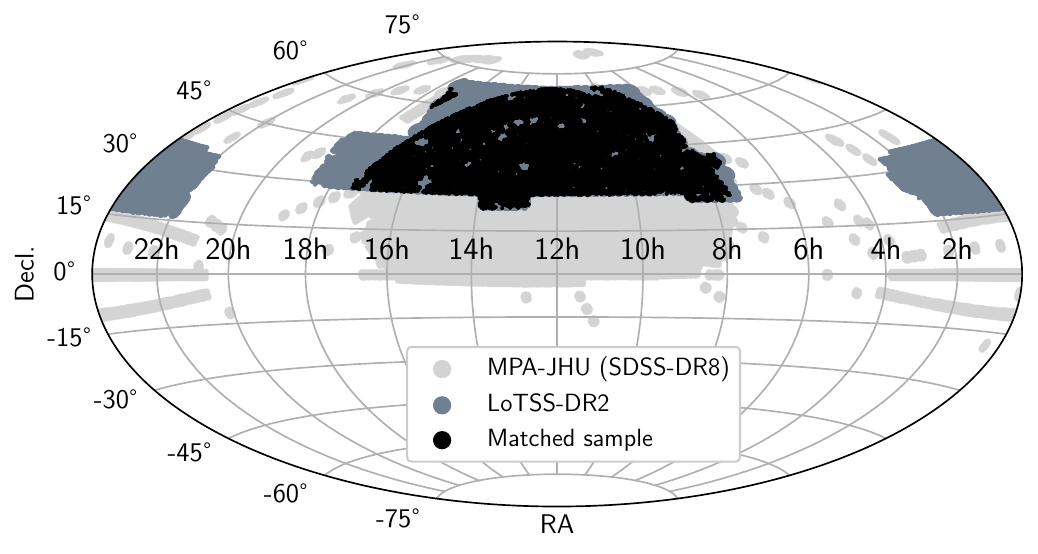}
    \caption{Sky coverage of MPA--JHU DR8 and LoTSS-DR2. The light and dark grey regions indicate sources from the MPA--JHU and LoTSS-DR2 catalogues respectively. The black points represent sources that belong to the parent sample as defined in Section \ref{selection}, i.e. sources that have optical as well as radio observations.}
    \label{fig:sky_coverage}
\end{figure}

We used 150\,MHz\footnote{The central frequency of the LoTSS data is at or near 144\,MHz, but we use 150\,MHz in this work for consistency with previous works.} LoTSS \citep{2017Shimwell}  observations, specifically from the second data release of the survey (LoTSS-DR2; \citealt{Shimwell22}). The survey covers 5634 square degrees of the northern sky, with 6-arcsecond resolution and a median root mean square (rms) sensitivity of 83\,$\mu$Jy per beam. DR2 contains not only the 150\,MHz images, RMS images and source catalogues in this area but also a value-added catalogue, which includes cross-identifications with optical counterparts from the DESI Legacy Imaging Surveys \citep{2019Dey_LEGACY} as described by \citet[hereafter \citetalias{2023MartinH_optcounterparts}]{2023MartinH_optcounterparts}, In brief, for large sources with extended radio emission, visual cross-identification and association were carried out through a citizen science project on Zooniverse\footnote{\url{http://lofargalaxyzoo.nl/}}. For compact sources, statistical techniques including but not limited to colour- and magnitude-dependent likelihood ratio methods \citep[e.g.][]{sutherland1992,ciliegi2003,smith2011} were used, as described by \cite{2019Wendy_optmatches}. \\
\\
The availability of the images alongside the catalogues allows us to perform forced aperture photometry for those objects in our sample which are not associated with 150\,MHz sources in the \citetalias{2023MartinH_optcounterparts} catalogue, as described in Section \ref{ap_phot}. The sky coverage of LoTSS-DR2 is shown in dark grey in Fig. \ref{fig:sky_coverage}.
\subsection{The MPA--JHU Spectroscopic Catalogue}
In this work we have made extensive use of the value-added catalogue of the eighth data release of Sloan Digital Sky Survey (SDSS) DR8 \citep{2011SDSSDR8} compiled by the group at Max Planck Institute for Astrophysics and the John Hopkins University (MPA--JHU; \citealt{Brinchman2004}). This catalogue includes flux measurements and uncertainties for a range of nebular emission lines including the H$\beta$, [O\textsc{iii}$]_{5007}$, H$\alpha$ and [N\textsc{ii}]$_{6584}$ lines commonly used for emission line source classification \citep{BPT1981,1987Veilleux}, as well as estimates of SFRs and stellar masses for each source. The sky coverage of the MPA--JHU catalogue, and the area in common with LoTSS-DR2 are shown in light grey and black (respectively) in Fig. \ref{fig:sky_coverage}.\\
\\
Within a three arcsec fiber aperture, galaxy SFRs are computed based on the nebular emission lines \citep{Brinchman2004}, using H$\alpha$ luminosity such that:
\begin{equation}\label{OptSFR}
    \textrm{SFR} (M_{\odot}\textrm{yr}^{-1})=  10^{-41.28} L_{\rm{H \alpha}} \textrm{ erg\ s$^{-1}$},
\end{equation}
\noindent where the authors have adopted an initial mass function (IMF) from  \citet{Kroupa2001}. Dust corrections were made using the H$\alpha$/H$\beta$ ratio and assuming a fixed unattenuated Case B ratio. Aperture corrections to correct for star formation not captured by the fibre measurement were estimated using photometry following \cite{Salim2007}. Since the spectra are measured through three arcsec apertures, where the fibre is positioned close to the galaxy's centre, it represents only 17--50 per cent of the entire galaxy's mass, thus potentially inducing aperture biases (e.g. \citealt{2016Richards_SAMIbiases}, \citealt{2017Duarte}). This means that the measured values might be appropriate for the galaxy's central bulge but not necessarily the outer discs, which can host significant star formation \citep{2003Kauffman}. To address this, we use aperture-corrected estimates of SFR and stellar mass from \citep{Brinchman2004}, who used resolved photometric colour information to empirically correct for aperture bias, and who updated the uncertainties on the aperture-corrected SFRs to account for the uncertain aperture-correction. \\ 
\\
In MPA--JHU DR8, the stellar mass values are estimated as described by \cite{2003Kauffman}, with the exception that $ugriz$ galaxy photometry alone is used rather than spectral indices Dn(4000) and H$\delta$ (which are obtained from the spectra). We use the total stellar masses, estimated using model magnitudes and photometry corrected for nebular emission using the spectra, and adopting a Kroupa IMF. In addition to the MPA--JHU SFRs and stellar masses for galaxies, we also considered their velocity dispersion and gas-phase nebular oxygen abundance values, measured using strong optical emission lines as described by \cite{2004Tremonti} and \cite{Brinchman2004}.

\subsection{The \textit{GALEX}-SDSS-\textit{WISE} Legacy Catalogue}

The \textit{GALEX}-SDSS-\textit{WISE} Legacy Catalog (GSWLC; \citealt{Salim2016}) combines \textit{GALEX} (UV; \citealt{Morrissey2007}) and optical data from SDSS DR4 \citep{Adelman2006} with Spectral Energy Distribution (SED) fitting done using CIGALE \citep{Noll2009}, assuming a Chabrier IMF \citep{Chabrier2003}. For consistency, the SFRs and stellar masses from the MPA--JHU catalogue were also converted to a Chabrier IMF by dividing by 1.06, following \citet{MadauDickinson14}. The SED fitting builds on \cite{Salim2007} by additionally including blending-corrected low resolution UV photometry, emission line corrections and flexible dust attenuation curves. We use the GSWLC-X2 catalogue, which covers 90 per cent of the SDSS sky coverage and also includes \textit{WISE} IR fluxes (at 12\,$\mu$m and 22\,$\mu$m) when calculating galaxy properties by applying energy-balance SED fitting. This technique assumes energy conservation between UV-optical stellar light and the light absorbed and re-remitted by dust in the infrared wavelengths. Because of the lack of data beyond 22\,$\mu$m, in GSWLC-X2 the IR and stellar emission were fit separately: IR luminosities were obtained directly from 22\,$\mu$m \textit{WISE} observations and then compared to SED-fit results of the stellar emission to verify energy balance. These luminosities were then calibrated against a \textit{Herschel} subsample with far-IR data, yielding a systematic offset of $ \leq \sim$0.01 dex for 22\,$\mu$m luminosities. Therefore, while individual IR luminosities may be very uncertain, the WISE 22\,$\mu$m–based estimates provide plausible average total IR luminosities  for statistical studies. By leveraging the energy-balance criterion in this way, the GSWLC-X2 catalogue is able to produce estimates of the current stellar mass and SFR averaged over the past 100\,Myr. For a more detailed discussion on the SED fitting procedure, we refer the reader to \citet{2018Salim}.

\section{Sample Definition}
\label{selection}

    In this work our focus is on SFGs, but it is well-known that AGN exist at a broad range of radio luminosities (e.g.  \citealt{2019aHardcastle}, \citealt{drake2024}). Therefore, to limit AGN contamination in our sample, we use an optical emission line diagnostic proposed by \cite{2006Kewley}, first described by \cite{BPT1981}. This is defined in the MPA--JHU catalogue as parameter \texttt{BPTCLASS} and we set \texttt{BPTCLASS} == 1 to obtain 215,224 SFGs. These galaxies were chosen to have a S/N > 3 on the BPT emission lines (H$\alpha$, [NII], [OIII], H$\beta$). Furthermore, following the recommendation of \citet{2005Kewley} we selected only those galaxies with $z\ge0.04$ to minimise the influence of the most unreliable aperture corrections, and this also limits the potential influence of sources with radio emission more extended than the 6 arcsec beam of LOFAR. We cross-matched this resulting sample of 167,274 galaxies with the \textit{GALEX}-SDSS-\textit{WISE} Legacy Catalogue using a nearest-neighbour method with a one arcsec search radius with \texttt{TOPCAT} \citep{2005TOPCAT} to obtain SED-fit parameters. 

Since both MPA--JHU and GSWLC adopt SDSS coordinates, 99.9 per cent of matched sources are within 0.1 arcsec, and changing the maximum search radius by $\pm 20$\,percent does not significantly impact our results. The GSWLC-X2 catalogue covers 90 per cent of the SDSS area, up to a redshift of 0.3, giving us matches to 114,519 sources between $0.04\,\leq\,z\,\leq\,0.3$. For consistency with the optical selection used in GSWLC and MPA--JHU catalogues, we use only the subset of sources selected in the SDSS Main Galaxy Survey galaxies which targets galaxies with photometry brighter than $r_{\rm{petro}}$\,= 17.77 \citep{2002Strauss_MGS}, reducing our sample to 111,292 sources.\\
\\
To select galaxies where radio emission is driven primarily by stellar processes \citep{Condon1992}, we restrict our sample to include only those sources with MPA--JHU and GSWLC stellar mass between  \textbf{$7.8\leq\log(M_*/M_{\odot})\leq12$}, and SFRs between $-3\leq$\,log($\psi/M_{\odot}\,\rm{yr}^{-1})\leq 3$, reducing the sample to 109,767 galaxies. By ensuring that the emission line measurements are flagged as \texttt{reliable}, we get 109,765 sources. We set the SDSS \texttt{targettype} and \texttt{spectrotype = `galaxy'} and consider only those galaxies where the redshift measurements are reliable by setting the flag \texttt{Z\_WARNING} to 0. This reduces our sample to 105,634 galaxies. We then select those galaxies observed as primary targets, and apply the cuts \texttt{V\_DISP $\sigma$ $>$ 0}\,km\,s$^{-1}$ and \texttt{OH\_P50 $>$ 0} to remove unphysical values of velocity dispersion and gas-phase metallicity respectively, reducing the sample to 97,846 galaxies. 
To make sure that the velocity dispersion fits are reliable, we apply the cut \texttt{V\_DISP\_ERR $>$ 0}\,km\,s$^{-1}$, which gives us 97,229 galaxies. Furthermore, for SDSS DR8, the template spectra used to calculate the velocity dispersion are convolved to a maximum value of $\sigma\sim420$\,km\,s$^{-1}$. Therefore, we remove galaxies with velocity dispersion values greater than 420\,km\,s$^{-1}$ as they might be unreliable\footnote{\url{https://www.sdss3.org/dr8/algorithms/veldisp.php}}, bringing the final sample with optical measurements to 97,141 galaxies. We then matched this parent sample with the LoTSS-DR2 catalogue, using a one arcsec\footnote{Varying the search radius by 20\% to 0.8 or 1.2 arcsec changes the number of matches by fewer than 50 sources ($<$0.5 per cent of our sample size) and has no significant impact on our results.} search radius and a nearest-neighbour algorithm using the SDSS positions from MPA--JHU and the positions of the cross-identified counterparts taken from the Legacy Surveys data used by \citetalias{2023MartinH_optcounterparts}. In this way, we obtained catalogue 150\,MHz flux densities for 25,449 galaxies. The choice of a one arcsec radius is motivated by the high astrometric precision of both catalogues, with the majority of positional offsets well described by a Gaussian distribution centred at $\sim$ 0.1 arcsec.
\subsection{150 MHz Aperture Photometry}\label{ap_phot}
To obtain radio flux densities for the sources not in the \citetalias{2023MartinH_optcounterparts} catalogue, (i.e. those which are below the PyBDSF\footnote{\url{https://pybdsf.readthedocs.io/en/latest/index.html}} detection threshold), we first identify the subset of the 97,141 sources located in the LoTSS-DR2 coverage by using a multi-order coverage (MOC) map. This subset contains 40,022 galaxies and is shaded black in Figure \ref{fig:sky_coverage} (of which 25,449 already have catalogued flux density measurements from \citetalias{2023MartinH_optcounterparts}). We perform aperture photometry on the LoTSS images by summing the flux densities within a ten arcsecond diameter circular aperture around the MPA--JHU positions, in a manner similar to \citetalias{Gurkan2018}, accounting for the resolution of the LOFAR maps. For galaxies in the \citetalias{2023MartinH_optcounterparts} catalogue, we verify that the flux densities measured using aperture photometry are consistent with those in the catalogue, with a typical scatter of $\sim$0.1 dex and median ratio of 0.974. Similarly, for 150\,MHz flux density uncertainties, we sum up the individual pixel variances in quadrature within the ten arcsec aperture from the LoTSS RMS maps. These uncertainties agree with those in the catalogue, with a scatter of $\sim$0.15 dex. We also verify that the RMS values are reasonable by measuring the aperture flux densities and RMS values at random sky positions with no radio sources and calculating the signal-to-noise ratios; the distribution obtained is similar to a standard normal distribution. For the rest of this paper, for sources with matches in the \citetalias{2023MartinH_optcounterparts} catalogue, we use the radio flux densities and uncertainties from the catalogue. For sources without matches in the \citetalias{2023MartinH_optcounterparts} catalogue, we use radio flux densities and uncertainties measured using aperture photometry.

\subsection{Radio luminosity}
To estimate the 150\,MHz luminosity for a source of flux density $S_{\nu}$ at frequency $\nu$ (whether from the \citetalias{2023MartinH_optcounterparts} catalogue or from our aperture measurements), we assume that radio sources obey a radio spectrum described by a simple power-law $S_{\nu} \propto \nu^{\alpha}$, where $\alpha$ is the spectral index. We perform the standard radio $K$-correction and get the rest-frame radio luminosity values $L_{\nu}$ to be:
\begin{equation}
    L_{\nu} = 4 \pi d_L^2 S_{\nu,\rm{obs}}(1+z)^{-\alpha- 1},
\end{equation}
\noindent where $d_L$ represents the luminosity distance at MPA--JHU spectroscopic redshift $z$ calculated in our adopted cosmology and the (1+$z$)$^{-1}$ accounts for bandwidth correction. We assume $\alpha=\,-0.7$, the typical median value for SFGs with shock-accelerated cosmic ray electrons (found by works such as \citealt{2016Hardcastle} and \citealt{2017Rivera} at 150\,MHz using LOFAR observations). Similarly, \citet{2018DeGasperin} show that between 147\,MHz and 1.4\,GHz the median spectral index is $\alpha\,\sim-0.78$ with a scatter of 0.24 dex, but that a flattening to $\alpha\,\sim-0.5$ occurs at the faintest flux densities and/or lowest frequencies. So using a constant spectral index value to estimate K-corrected luminosity densities could be an additional source of uncertainty, with previous works having observed an rms scatter of $\sim$ 0.38 on median spectral index values \citep{2013Mauch}, but as noted by \citet{2018SRead}, the additional uncertainty would be very small for local SFGs ($z\leq 0.3$) such as our sample. Therefore, while well‐motivated for “typical” star‐forming galaxies, some bias may persist due to spectral curvature, absorption or core‐dominated spectra. In principle, once radio surveys with depth and resolution comparable to LoTSS are available at other frequencies, more precise galaxy-by-galaxy spectral indices could be derived and applied. For the present study, however, such data are not yet available for the full sample, and we therefore adopt the canonical value of $\alpha= -0.7$ for consistency with the literature.

\begin{figure*}
    \centering
    \includegraphics[scale=0.26]{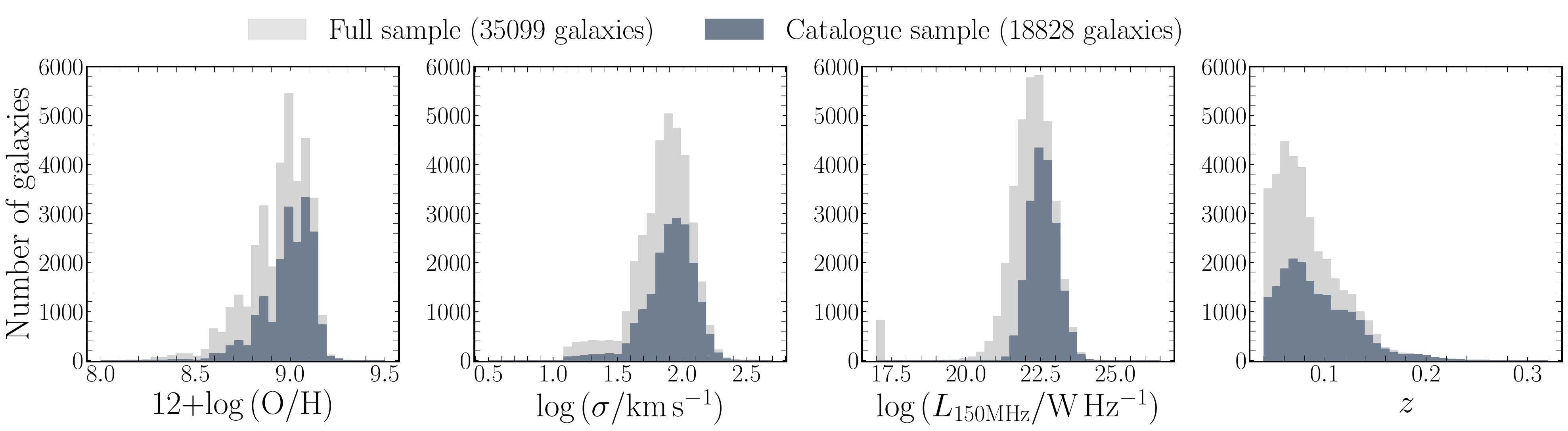}
    \caption{Distribution of gas phase metallicity, velocity dispersion, radio luminosity and redshift for our samples. The 18,828 galaxies which contain LoTSS flux densities given in the \citetalias{2023MartinH_optcounterparts} catalogue are shown in dark grey. For the sample of 35,099 galaxies (18,828 sources with catalogued radio flux densities and aperture flux densities measured for remaining 16,271 sources), the distributions are shown in light grey.} 
    \label{fig:histograms}
\end{figure*}

\begin{figure*}
     \begin{minipage}{0.5\textwidth}
       \centering
       \includegraphics[width=1.\linewidth]{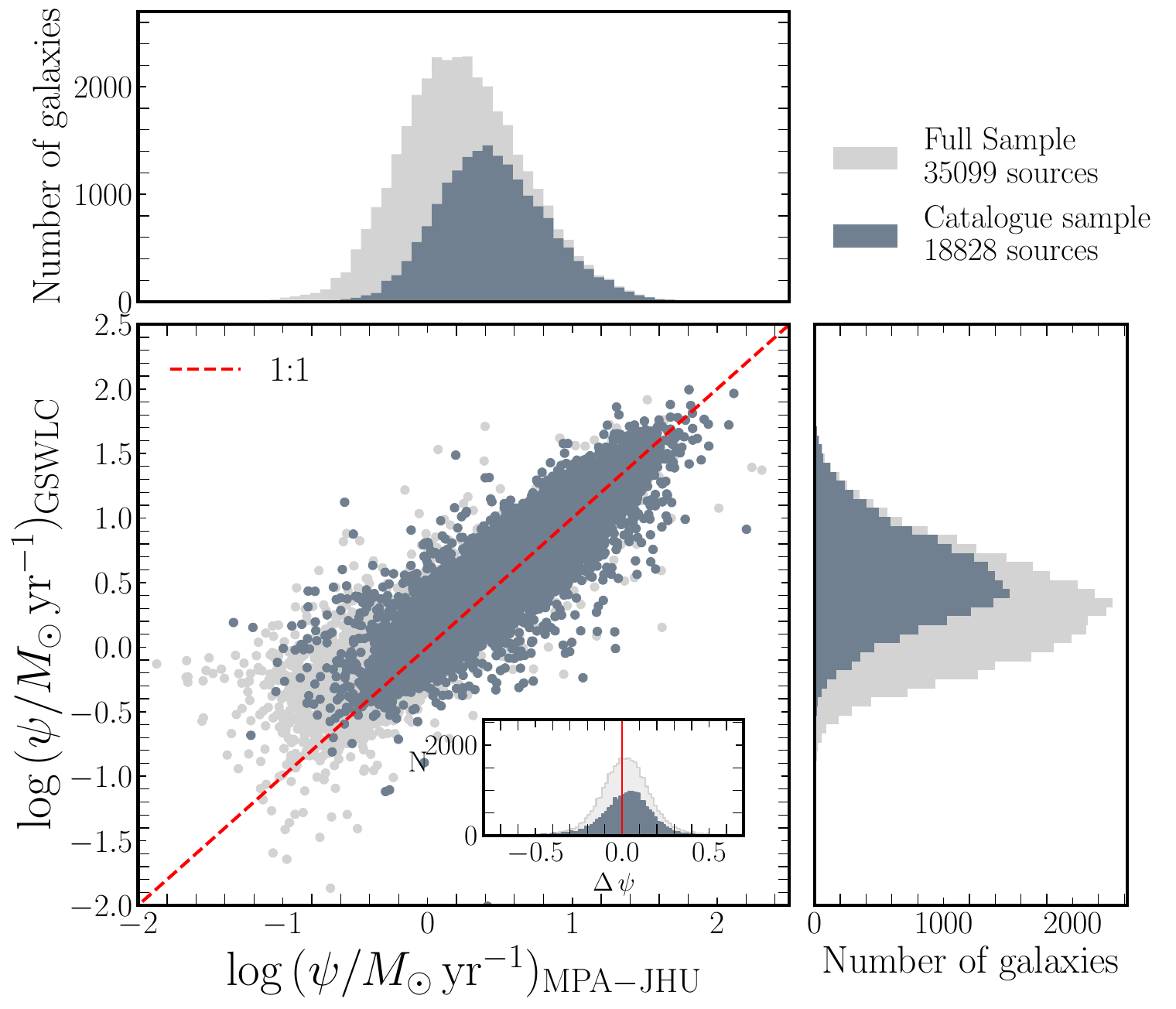}
     \end{minipage}\hfill
     \begin{minipage}{0.5\textwidth}
     \centering
       \includegraphics[width=1.\linewidth]{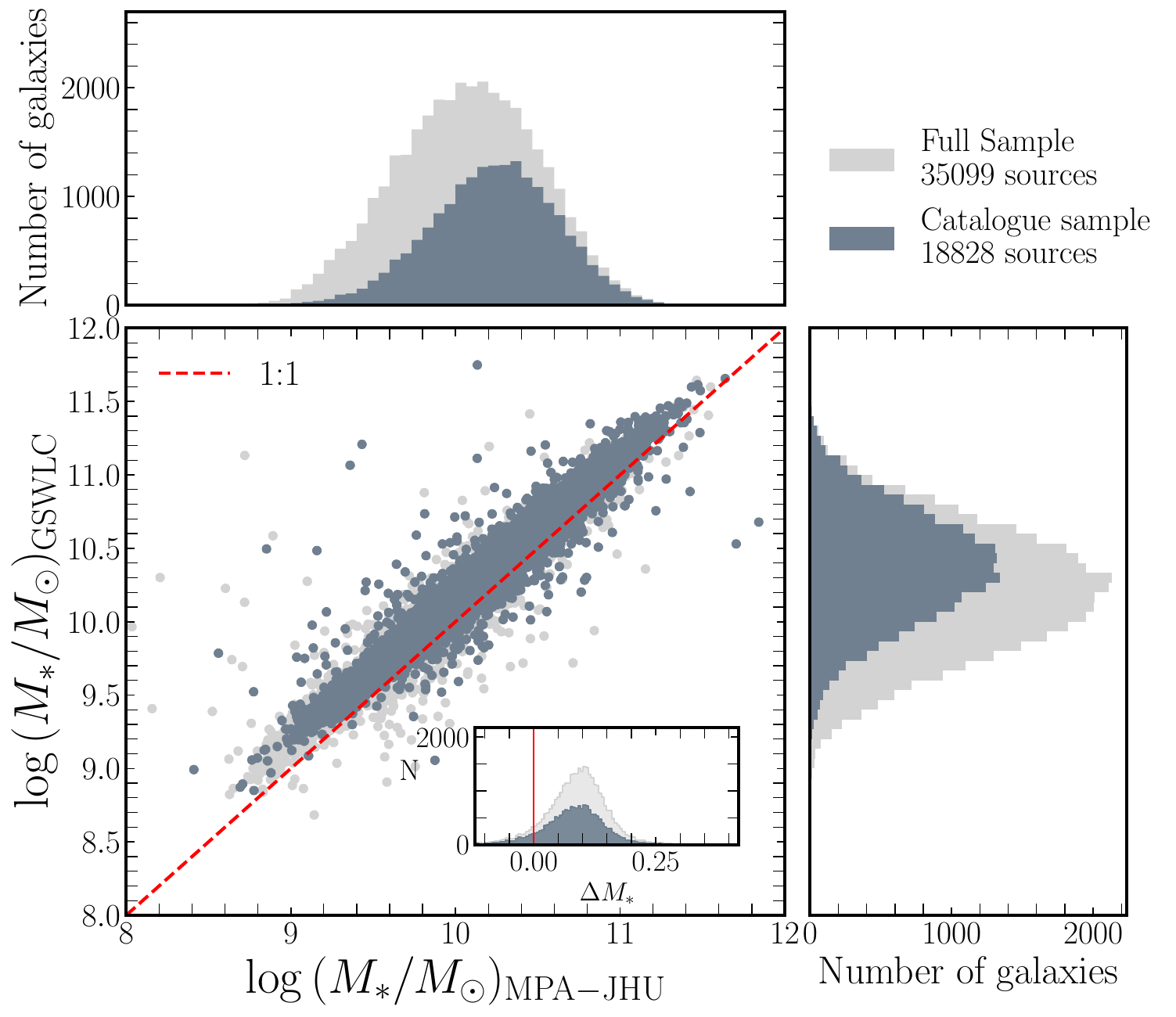}
     \end{minipage}
    \caption{Comparison of SFR and stellar mass values derived using the photometric (GSWLC-X2) and spectroscopic (MPA--JHU) datasets for sources with 150\,MHz flux densities in the LoTSS catalogue (in dark grey) and sources with radio luminosities measured with aperture photometry (in light grey). The inset plot shows the difference between SFR and stellar mass, and all values have been converted to our adopted Chabrier IMF (where required).}\label{fig:sfrmasscomparisonplot}
\end{figure*}

\begin{figure}
    \centering
    \includegraphics[ width=0.99\columnwidth]{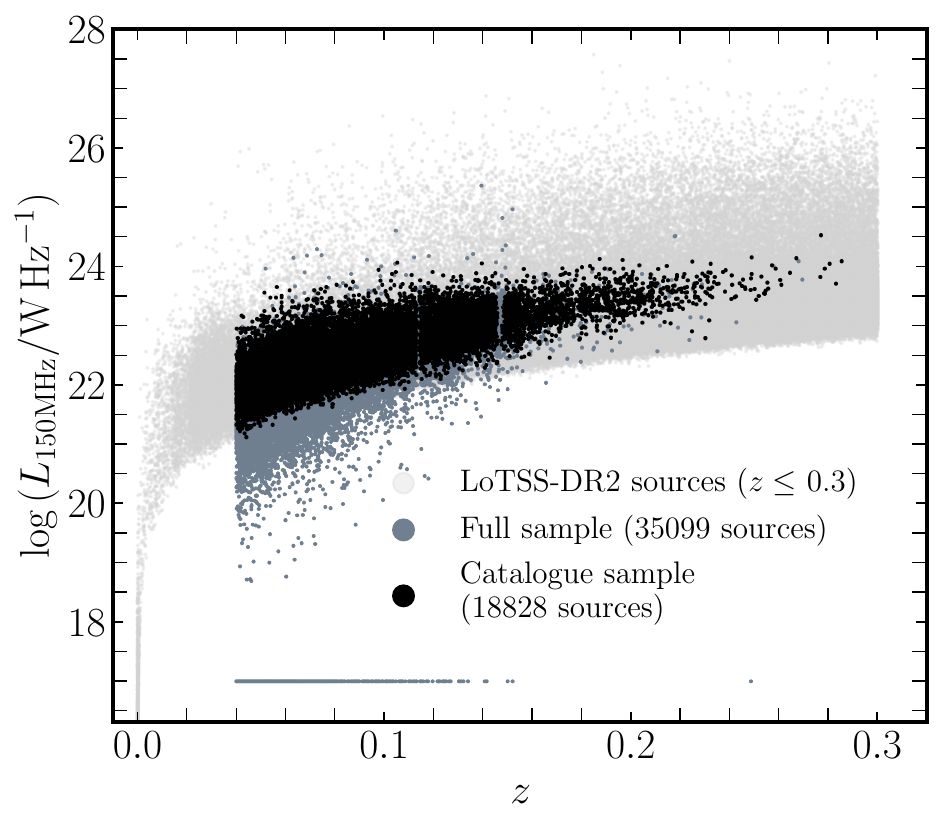}
    \caption{The relationship between redshift and 150\,MHz luminosity for our samples. All sources in the LoTSS-DR2 catalogue with redshifts $z \leq 0.3$ are shown in light grey, while the subset of sources with LoTSS-DR2 catalogue matches in our spectroscopic sample are shown in black. These have been complemented by the sources for which we have measured aperture 150\,MHz flux densities, which are shown in dark grey (as indicated in the legend). } 
    \label{fig:bpt_lumdis}
\end{figure}

\subsection{Minimising residual AGN contamination}
To reduce the possible influence of optically thick AGN, which may not be detected via the BPT diagnostic but can still show strong radio emission, we use two further diagnostics. Firstly, we use the mid-IR selection criteria from \citet{2018Assef} to exclude possible AGN based on their \textit{WISE} colours. These sources are identified using:

\begin{enumerate}
\itemsep-0.3em
    \item If $W2$  $> \gamma_R$, then $W1 - W2$  $> \alpha_R$ exp[$\beta_R$ ($W2$ - $\gamma_R$)$^2$]
    \item If $W2$ < $\gamma_R$, then $W1-W2$ > $\alpha_R$
\end{enumerate}
\noindent where $W1$ and $W2$ are magnitudes in the \textit{WISE} bands centered at 3.4 and 4.6 microns respectively, values taken from \citetalias{2023MartinH_optcounterparts}, and ($\alpha_R$, $\beta_R$, $\gamma_R$) = (0.662, 0.232, 13.97), and these criteria are expected to be 90 per cent reliable. This returned three AGN candidates, which were excluded from the parent sample, giving us 40,019 galaxies in total with 25,446 galaxies of those present in \citetalias{2023MartinH_optcounterparts}. \\
\\
Secondly, we need to remove galaxies with a radio luminosity higher than can be expected on the basis of their H$\alpha$ luminosity (i.e. those with a radio excess), likely indicative of AGN activity. To identify such sources, we cross-match our sample with the sample given by \cite{drake2024}, who performed a probabilistic spectroscopic classification on the LoTSS-DR2 radio-selected sources. We match the samples using their radio \texttt{Source\_Name} to obtain 25,034 sources (out of the 25,446 galaxies, 412 galaxies are removed from the sample either due to size cuts or unreliable fits of prominent BPT emission lines like H$\beta$ and [OIII]$_{5007}$ lines where they coincide with sky lines, flagged as \texttt{FIT\_WARNING} in the SDSS value-added catalogue by the Portsmouth group \citep{2013Thomas_Portsmouthcatalogue} which was used for the probabilistic classifications). We note that this is also evident in Figure \ref{fig:bpt_lumdis} as regions with fewer data points, see e.g. $z \sim 0.15$. 550 sources were removed due to radio excess cuts but we repeat our analysis with these sources included and find that our results remain consistent within statistical uncertainties.\\
\\
We set the criterion \texttt{CLASS\_SFG\,$>$\,0.9} from \citet{drake2024}, selecting galaxies classified as star-forming with at least 90 per cent confidence. This threshold not only makes sure that the source falls in the SFG part of the [NII]-BPT diagram but also ensures no radio excess is measured. In doing so, our catalogue sample reduces to 19,757 sources. Following  the recommendations of \citet{drake2024}, we consider only sources with \texttt{zscore\,<\,2.5}\footnote{Metric defined in \cite{drake2024} to quantify the significance of deviation from the null hypothesis that the classification is accurate}, reducing the sample to 19,751 galaxies. We then select those galaxies that are in \citetalias{2023MartinH_optcounterparts} with a S/N $\geq$ 3 on total radio continuum flux density, and obtain 18,828 galaxies; we call this the `catalogue sample'. Similarly, for the `full sample', i.e. including sources with aperture flux densities in addition to those with flux densities from the \citetalias{2023MartinH_optcounterparts} catalogue, we obtain 35,099 galaxies  out of 40,019 galaxies after removing galaxies with \texttt{CLASS\_SFG\,<=\,0.9} as given in \citet{drake2024}. \\ 
\\
Thus, we compile a `full sample' of 35,099 BPT-classified star-forming galaxies, with catalogue radio flux density measurements where available and aperture 150\,MHz flux densities otherwise. This sample will be employed in Section \ref{rcsfrmass} to determine the RC–SFR relation. Within it, the `catalogue sample' comprises 18,828 galaxies with a signal-to-noise ratio (S/N) $\geq$ 3 in 150\,MHz flux density from \citetalias{2023MartinH_optcounterparts}. This subset will be analysed using a Random Forest algorithm to predict radio luminosities and identify the most important galaxy features driving these predictions.

\subsection{Sample properties}

The distributions of gas phase metallicity, velocity dispersion, radio luminosity and redshift for our samples are shown in Figure \ref{fig:histograms}, with the distributions of SFR and stellar mass estimates (from both MPA--JHU and GSWLC) shown in marginal histograms in the left and right panels of Figure \ref{fig:sfrmasscomparisonplot}. In each of these figures, the catalogue sample is shown in dark grey, and the full sample is shown in light grey. \\
\\
In the main panels of Figure \ref{fig:sfrmasscomparisonplot}, we compare the MPA--JHU and GSWLC SFRs (left) and stellar masses (right), after converting the MPA--JHU values to our adopted Chabrier IMF by dividing by 1.06 following \citet{MadauDickinson14}. In the left panel, there is good agreement between SFRs reported in MPA--JHU DR8 and GSWLC catalogues, with a scatter of $\sim$0.2 dex. In the right panel of Figure \ref{fig:sfrmasscomparisonplot} we see that stellar mass values from GSWLC are on average around 0.1\,dex  higher compared to the ones reported in MPA--JHU with $\sim$0.09 dex scatter if we compare the two (as shown in the inset). This scatter is comparable to the values reported by \citet{Salim2016}, who attributed this difference in stellar mass values to different star-formation history assumptions and the inclusion of UV photometry in GSWLC-X2.\\ %
\\
Lastly, in Figure \ref{fig:bpt_lumdis} we show the distribution of radio luminosity log $L_{150\rm{\,MHz}}$ with redshift $z$. The catalogue sample is shown in black, and the full sample is shown in dark grey. As expected, and discussed in \citetalias{Smith20}, some of these sources have unphysical values of flux densities; following \citetalias{Smith20}, sources with $\log (L_{150\rm{\,MHz}}/$W\,Hz$^{-1}$)\,<$\,17$ are arbitrarily assigned $\log(L_{150\rm{\,MHz}}/$W\,Hz$^{-1}$)\,=$\,17$ and included in the sample; these sources are visible as the horizontal light grey stripe in Figure \ref{fig:bpt_lumdis}, and constitute $\sim2\%$ of our full sample. As shown in the figure, the use of an optically selected sample results in the exclusion of many faint radio sources. This limitation will be addressed by the upcoming WEAVE-LOFAR survey \citep{2016Smithweave}, which is expected to obtain more than $10^6$ optical spectra for LOFAR-detected radio sources.

\section{Method and Results}\label{results}

In this work, we aim to determine the RC--SFR relation with unprecedented accuracy. We will begin by determining which parameters are important in predicting a SFG's radio luminosity (in section \ref{ml}), before building on those results to fit for the analytic form of the RC--SFR relation.

\subsection{Which galaxy parameters matter?}\label{ml}

To reveal the \textit{importance} of different galaxy parameters in predicting a SFG's radio luminosity in a non-parametric manner, we use the Random Forest ensemble learning method \citep{2001breiman}. We generate a non-parametric model to predict radio luminosity values of our catalogue sample of SFGs using five different combinations of parameters:

\begin{enumerate}
    \item [$\blacksquare$] \textbf{Model 1 -} Spectroscopic: both SFR and stellar mass estimated from optical spectroscopy, taken from the MPA--JHU catalogue.
     \item [$\blacksquare$] \textbf{Model 2 -} Spectroscopic $\boldsymbol{+}$ (O/H) \& $\sigma$ : Along with SFR and stellar mass estimates from MPA--JHU, we also include gas phase metallicity and velocity dispersion in the model.
    \item [$\blacksquare$] \textbf{Model 3 -} Photometric: both SFR and stellar mass estimates come from the GSWLC SED-fitting of broadband photometry including UV, optical and infrared observations.
    \end{enumerate}
Additionally, we examine the stellar mass dependence of the relation when we interchange the SFRs and stellar mass values derived from the spectroscopic and photometric datasets, i.e. 
\begin{enumerate}
    \setcounter{enumi}{3}
    \item [$\blacksquare$] \textbf{Model 4 -} Photo-Spec: SED-fit SFR from GSWLC and stellar mass values from the MPA--JHU catalogue.
    \item [$\blacksquare$] \textbf{Model 5 -} Spec-Photo: Balmer-line estimated (aperture-corrected) SFR from MPA--JHU and stellar mass from broadband photometry SED-fitting (GSWLC).
\end{enumerate}

\subsubsection{Random Forest Regression}

We perform a machine learning analysis using the \texttt{scikit-learn} package \citep{scikit-learn}. A Random \textit{forest} regressor, as the name suggests, is a random combination of multiple decision \textit{trees}. In this context, a decision tree is a supervised non-parametric machine learning algorithm (i.e. the target label or class to be predicted is already known, as opposed to unsupervised techniques where the machine looks for patterns in a dataset without being explicitly given output labels; \citealt{Breiman1984ClassificationAR}). A decision tree uses a flowchart-like structure with \textit{recursive binary splitting}: it starts with all the observations from the top of the tree, dividing the predictor space into two new branches in each iteration down the tree \citep{Strobl_recursivebsplit}. It is a \textit{top-down greedy} method because the algorithm looks for the best available variable in only the current split with no concern about future splits that lead to a superior tree. Therefore, an advantage of RF analysis is that it allows for an explicit calculation of the \textit{relative importance} of each input galaxy feature in predicting the target value by assigning weights to the features at each split. In our case, it would be the relative importance of SFR, stellar mass, gas phase metallicity and velocity dispersion in predicting radio luminosity.

\subsubsection{Cross-validation and prevention of overfitting}\label{kfold}

In supervised machine-learning techniques, cross-validation is standard practice to build and evaluate a model \citep{2001CrossVal}. We do this by using three disjoint subsets: a training set to learn the hidden correlation between input features and the output label, a validation set to measure the learning level and verify the absence of overfitting, and a final test set used to evaluate the overall performance of the trained and tested model using predetermined criteria (we used least mean-squared error). In this work, we have used 60 per cent of the data for training and 20 per cent each for the validation and test sets. \\
\\
Since the feature selection is randomised to prevent overfitting in a RF Regressor, cross-validation does not affect the outcomes of individual decision trees. Our main results also do not change with or without cross-validation as long as we control for overfitting, which can be checked in two ways: (a) when a model shows very low mean-squared error (MSE) on the training set predictions but a higher error on the test and validation sets, and (b) when the feature importance is different across the training and test or validation sets. This generally happens when the model memorizes the noisy data instead of the underlying relation. One way to control for overfitting is by increasing the number of galaxies in a \textit{leaf} node\footnote{The terminal node of the RF decision tree}, also known as an \textit{early stopping routine}. After setting the minimum number of samples in the leaf node to $\sim$70, selected based on cross-validation, our model has a prediction mean-squared error (MSE) of $\sim$ 0.12 $\pm$ 0.03 on all three subsets, indicating that the model does not overfit the data.

\subsubsection{Feature importance}\label{feature}

The purpose of including galaxy properties in this work is to establish which parameters can most accurately predict radio luminosities. A single decision tree assigns more weight or importance to those variables that can predict radio luminosity with the most accuracy at each split. It then adds up the weights to calculate the relative importance. In a random forest, the relative importance is averaged for all the $N$ number of trees (in our models, we have between $300-400$ trees, depending on the model). Since the model treats the input galaxy parameters as competitors and, at each decision fork, selects the one that minimizes a chosen metric (in our case, root mean-squared error or RMSE), we must note that the importance is only \textit{relative} because it depends on the input set of parameters.\\
\\
Feature importance values give us a general idea about which galaxy parameter is highly related to the target variable. However, we can not always be sure that the models that give the best predictions (least RMSE in this case) also give the most precise interpretation of the underlying causative processes. This is because multiple model instances can all give similar predictive results but have different functional forms. To understand this better and to check how much the RF model relies on each parameter for its predictions, we adopt a permuted feature importance (PFI) analysis described in detail by \citealt{2018FisherRudinDominici}. In the PFI technique, the model is re-run with the values of one feature randomly shuffled per run, breaking the link between that feature and the target radio luminosity for that run and the increase in model RMSE is estimated. The higher the increase in the RMSE value for a given feature, the more important the feature is for the model, as it relies more on that feature for accurate prediction. Similarly, a galaxy feature is not as crucial if randomly shuffling its values causes minimal or no change in the RMSE of the model predictions because, in this case, the model ignored that feature for the luminosity prediction. However, like the RF feature importance, the importance of the permuted feature does not show a parameter's intrinsic predictive value but only how important the parameter is for the specific model at hand. 
\begin{figure*}
       \centering
      \includegraphics[scale=0.5]{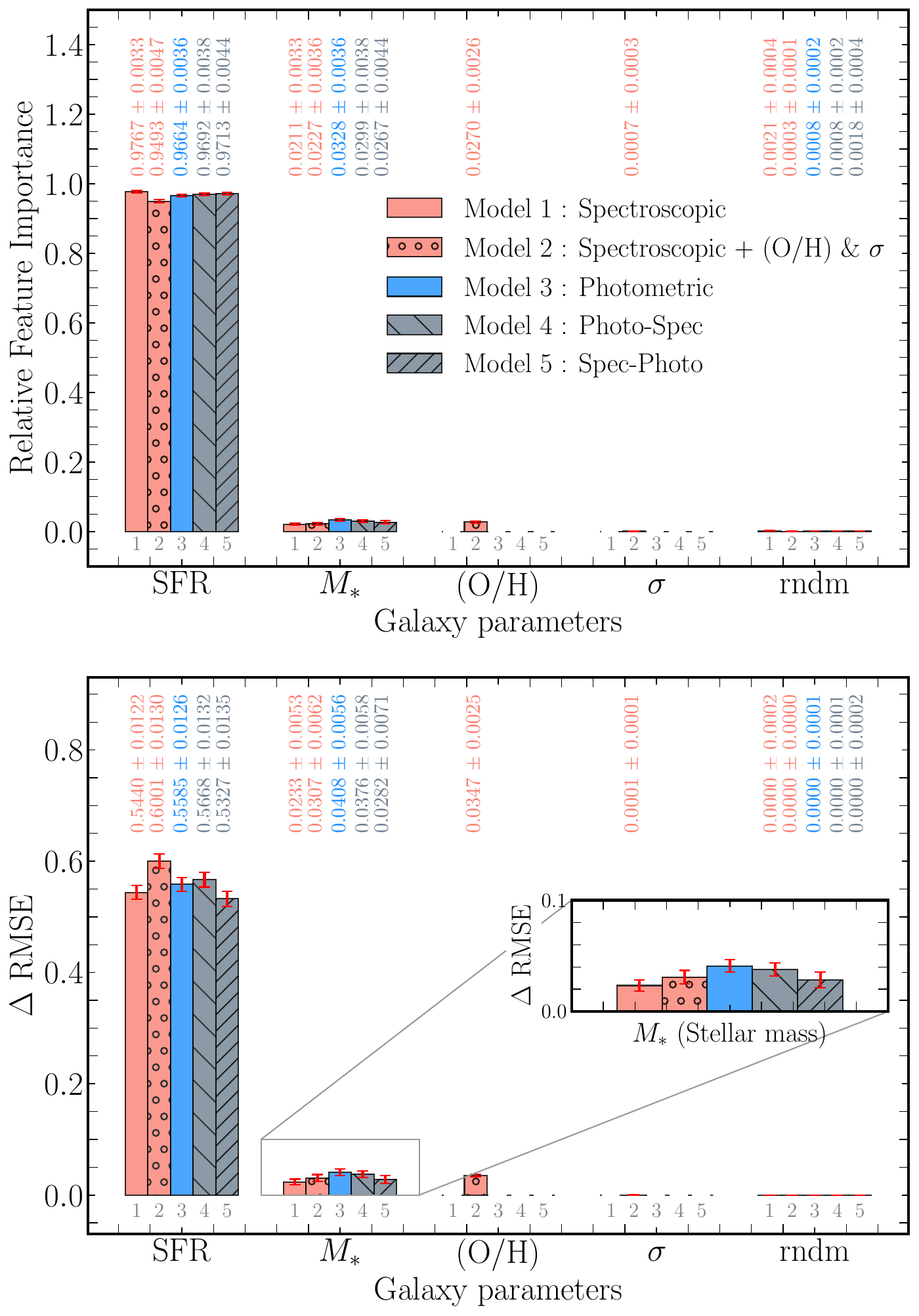}
      \caption{\textit{Upper panel:} relative feature importance of SFR, stellar mass, gas-phase metallicity and velocity dispersion in predicting radio luminosities for our sample of SFGs, when included in the model. The importance values of parameters in each model sum to 1. \textit{Lower panel:} permuted relative importance for the same features. The numbers below each bar correspond to the legend, and to the enumerated explanations in section \ref{ml}. } 
       \label{fig:RF_plot}
   \end{figure*}

\subsubsection{Random Forest model results}\label{rfresults}

We first \textit{robust} scale all the input features -- i.e. we subtract the median from each data point and normalise it by the interquartile range. Thus, all input data are in a unit-less format. This method is more robust to outliers compared to standard scaling where we perform mean subtraction and divide by the standard deviation. We note that our results remain consistent regardless of the scaling method applied. Additionally, in our case, running the RF regression algorithm even with unscaled values yields qualitatively similar results because the splits are determined by ordering the data (not by absolute values). We run the RF regressor with cross-validation as described in Section \ref{kfold} and set an early stopping routine, i.e. set the minimum number of samples in the leaf node (final split) to prevent overfitting. The list of features for all five models also includes a \texttt{rndm} parameter generated using the pseudo-random number generator of the \texttt{numpy.random}\footnote{\url{https://github.com/bashtage/randomgen}} module. Any feature with importance close to that of this parameter can be regarded as not important in the prediction of radio luminosity using that specific model. \\
\\
The feature importance (as described in Section \ref{feature}) of the different parameters is shown in the upper panel of Fig. \ref{fig:RF_plot} and the change in RMSE when each feature is permuted is shown in the lower panel of the same figure. The final feature importance is given by the median of the 100 runs, with a different \texttt{random\_state} in each run, and the standard deviation is taken as the uncertainty value. Reassuringly, the most important feature in every run is the SFR, whether it comes from the MPA--JHU or GSWLC catalogues. Furthermore, in every case, the stellar mass estimates are determined to have significant importance (i.e. the significance of the difference between the importance of stellar mass and that of the random variable is >\,5$\sigma$ for all models). This holds true irrespective of the use of photometric or spectroscopic data. Some authors also use Gradient boosting algorithms in place of RF Regression given that the former is able to handle missing data better. However, we find that our results do not change when we use \texttt{XGBoost} regressor \citep{xgboost} to predict radio luminosity. Similarly, we could also omit one feature at a time to evaluate its contribution in improving the MSE of the model predictions. In doing so, we find that the results are consistent within uncertainties with that of the permuted feature importance, for both RF regressor and \texttt{XGBoost} regressor model. More specifically, in both cases, SFR and stellar mass are still the most important features in predicting radio luminosity, in that order. We note that these results are robust to changes in IMF assumptions. \\
\\
Interestingly, we also observe significant importance of gas-phase metallicity in the spectroscopic$\boldsymbol{+}$ (O/H) \& $\sigma$ model. However, it is difficult to identify if the trend is real or an artefact, given that stellar mass and gas-phase metallicity values are well-known to be correlated through the mass-metallicity relation (e.g. \citealt{2004Tremonti}) and that they are both correlated with the SFR via the fundamental metallicity relation (FMR; e.g. \citealt{2010Mannucci}).  Permuted and drop column feature importance values are also known to be affected by correlated features, i.e. if two features are strongly correlated, the model still has access to the correlated feature when a feature is permuted or dropped (see e.g. \citealt{2008strobl}; \citealt{2009nicodemusPCC}). To expand on this further, we computed the partial Spearman rank correlation coefficients (e.g. \citealt{2005brown}) between spectroscopy-derived SFR and radio luminosity in two cases: controlling for (i) stellar mass and (ii) gas-phase metallicity. Controlling for stellar mass reduced the correlation from $\rho_{\rm{SFR, \mathnormal{L}}_{150}}$ = 0.862 ($p$-value $<$ 0.01) to $\rho_{\rm{SFR, \mathnormal{L}}_{150}\cdot \mathnormal{M_*}}$\,=\,0.675 ($p$-value $<$ 0.01), while controlling for metallicity had a much smaller effect ($\rho_{\rm{SFR, \mathnormal{L}}_{150}\cdot \rm{({O/H})}}$ = 0.848;  $p$-value $<$ 0.01), suggesting that stellar mass is a more significant confounding variable than metallicity.  In the absence of an obvious explanation for why gas-phase metallicity may be of importance for predicting a galaxy's radio luminosity, we therefore note that its apparent importance is of potential interest, but defer a more detailed analysis to a future work. The difference between correlation and causation is pertinent here, and this has been studied in some detail in the context of RF algorithms by \citet[][see appendix B.2 of that work]{Bluck2019}. \\
\\
Furthermore, whilst RF is robust against global measurement uncertainty, it can be influenced by differential measurement uncertainty i.e. the case where different features of the data have different noise properties -- as is the case in our dataset. Higher measurement uncertainty can also decrease the feature importance of a variable (see  e.g. appendix B.4 of \citealt{2022Piotrowska}). In our spectroscopic dataset, the uncertainties on gas-phase metallicity ($\bar{\sigma}_{(\rm{O/H})} \approx 0.03\,$dex) are significantly smaller than the uncertainties on stellar mass values ($\bar{\sigma}_{M_*} \approx 0.08\,$dex), which could partially inflate the importance of metallicity. In a similar manner, if we re-run the RF regression with the SFRs and stellar mass estimates from both MPA--JHU and GSWLC simultaneously, the model indicates an importance around 0.9 for the photometrically-derived SFRs, and around 0.1 for the spectroscopy-derived SFRs. This is on the one hand reassuring, given that the time-scales associated with using the UV continuum and radio continuum luminosity as SFR indicators are more comparable than that of the Balmer lines \citep[e.g.][]{1998AKennicutt_SF}; but on the other hand, it is hard to interpret given the heteroscedasticity, i.e. that the uncertainties on the SED-fit stellar mass and SFR estimates are around half the size of those from MPA--JHU.\\
\\
We also ran the regression models with morphological parameters, available reliably for 18,685 out of the 18,828 sources in our sample. The morphological parameters were taken from \citealt{2011Simard}, who performed 2D bulge+disc decompositions in the \textit{g} and \textit{r} bands for galaxies in the seventh data release of Sloan Digital Sky Survey (SDSS DR7; \citealt{2009Abazajian}). However, we find that importance of morphological parameters in predicting radio luminosity are consistent with zero once uncertainties are taken into account (see Appendix \ref{morphappendix}). We note, however, that the sample is restricted to star-forming galaxies, which preferentially selects disc-like systems. As a result, the lower importance of morphological parameters such as Sérsic index, in our analysis could indicate that we have fewer star-forming ellipticals rather than an absence of physical relevance. In a more morphologically diverse population, particularly one including early-type star-forming galaxies, morphology may play a more prominent role. For the rest of this paper, we will focus primarily on SFR and stellar mass, as described in the following section.

\begin{figure*}
    \centering
     \includegraphics[scale=0.67]{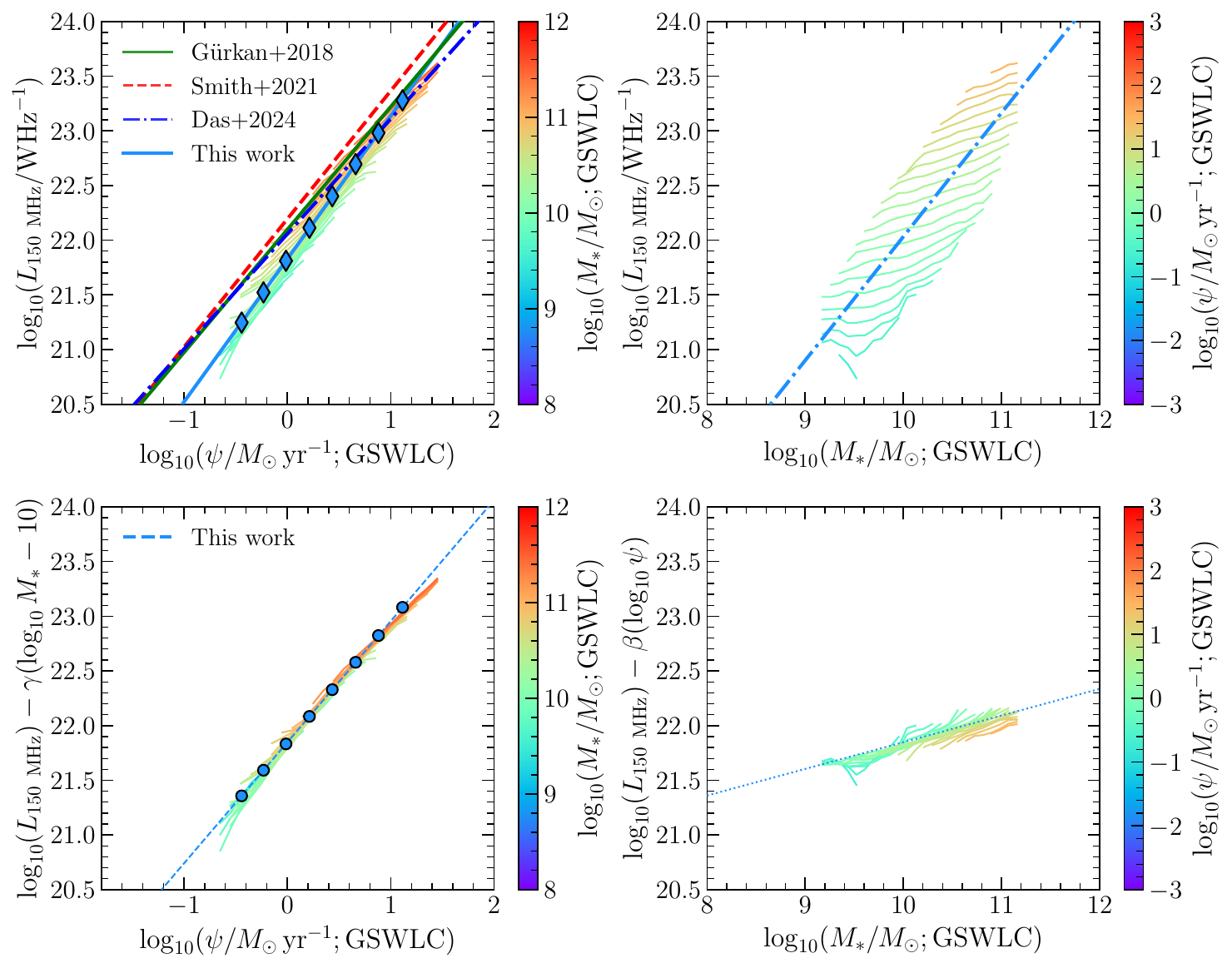}
     \caption{Dependence of radio luminosity at 150\,MHz on SFR and stellar mass. The top \textit{left} panel shows the radio luminosity at 150\,MHz as a function of photometrically-derived (i.e. GSWLC) star-formation rates. Each coloured line represents the median likelihood value of radio luminosity at a given SFR in stellar mass bins indicated by the colour bar. The mass-independent best-fitting line(s) from \citetalias{Gurkan2018} (in green), \citetalias{Smith20} (in red, dashed) and \citetalias{das2024} (in blue, dashed-dot) are overplotted. Top \textit{right} panel: Similar to the top left panel, but the median likelihood radio luminosity as a function of stellar mass are plotted in bins of SFR as indicated by the colour bar. The bottom left panel shows the RC--SFR relationship with the stellar mass dependence taken out using equation \ref{L150SFRm_equation} and the best-fit parameters obtained in Section \ref{rcsfrmass}. The bottom right panel shows the median likelihood radio luminosity as a function of stellar mass with SFR dependence taken out using equation \ref{L150SFRm_equation}. The radio luminosity values, estimated in bins of star formation rates (SFRs) using the mass-dependent form of the RC--SFR relation (given by equation \ref{lsfrm_eq_gswlc}), are represented by the diamond-shaped points in the top left panel. These values are calculated based on the median SFR and stellar mass within each SFR bin. The solid line represents the best-fit line through the diamond points. In the bottom left panel, we plot the radio luminosity values, shown as circles, calculated using median SFR values in each bin, without accounting for the stellar mass dependence.The dashed line represents the best-fit line through the circle points.}
     \label{fig:rcsfrmass_gswlc}
 \end{figure*}
 \begin{figure*}
    \centering
     \includegraphics[scale=0.67]{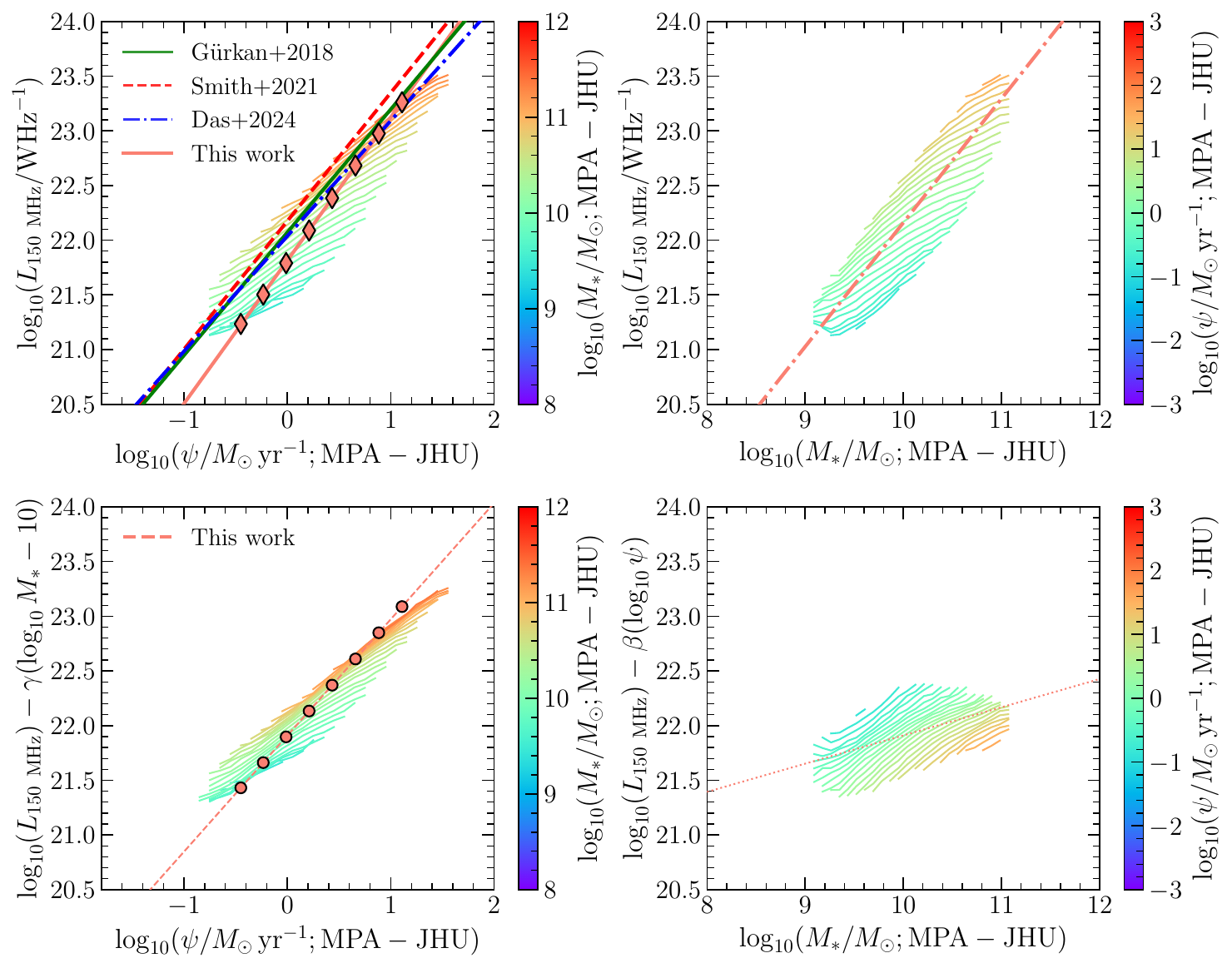}
     \caption{Same as Figure \ref{fig:rcsfrmass_gswlc} but with stellar mass and SFR values taken from the spectroscopic dataset (i.e. MPA--JHU).}
     \label{fig:rcsfr_mass_MPA--JHU}
 \end{figure*} 

\subsection{Stellar mass dependence of RC--SFR}\label{rcsfrmass}
Having demonstrated that the stellar mass is important for studying the relationship between a galaxy's SFR and 150\,MHz luminosity, we now quantify the stellar mass dependence of the RC--SFR relation. To do this, we follow the method of \citetalias{Smith20}, and calculate the median 150\,MHz luminosity as a function of SFR and stellar mass in three dimensions. We summarize the method as follows: to robustly account for our knowledge of the SFR, stellar mass and $L_{\rm{150\,MHz}}$ of each source in our sample, we build a three-dimensional PDF for each source, by generating 100 samples in each of the SFR, stellar mass (in logarithmic space) and radio luminosity (in linear space), in accordance with \citetalias{Smith20}. We do this for both spectroscopic as well as photometric data. For the sampling, we assume asymmetric uncertainties on SFRs and stellar masses by using the 16th, 50th and 84th percentiles provided in the MPA--JHU DR8 catalogue and for the photometrically-derived SFRs and stellar masses, we assume that the 50th$-$16th percentile and 84th$-$50th percentile are equal to the symmetric 1$\sigma$ uncertainty quoted in the GSWLC catalogue. We then bin the samples using fifty uniformly-spaced bins of stellar mass between $ 7<\rm{log}\,(\mathnormal{M}_* \slash \mathnormal{M}{_\odot})<12$, sixty bins for SFRs between $-3<\rm{log}(\psi\slash \mathnormal{M}{_\odot}\,yr^{-1})<3$ and 180 uniformly spaced bins for radio luminosities between  $ 17<\rm{log} (\mathnormal{L}_{\rm{150}}\slash\,W\,Hz ^{-1})<26$. We combine our knowledge of each source by adding each of these three-dimensional PDFs together. In each bin of SFR and stellar mass, we then calculate the median likelihood radio luminosity; these values are shown as a function of SFR and stellar mass in the top panels of Figure \ref{fig:rcsfrmass_gswlc} for photometrically-derived (i.e. GSWLC) SFRs and stellar mass estimates (on the left and right sides, respectively). In the left panels of the figure, the median likelihood lines are coloured by the GSWLC stellar mass in each bin. In the right panels, they are coloured by photometrically-derived SFRs. To limit the effect of low number statistics, we show only those bins which contain at least 15 galaxies (taking into account that each galaxy is sampled 100 times). Figure \ref{fig:rcsfr_mass_MPA--JHU} shows the corresponding values when the spectroscopic (i.e. MPA--JHU) SFRs and stellar mass estimates are used. Due to the different error budget in each catalogue, the scatter on the RC--SFR is much lower when GSWLC SED-fit SFRs are used (Figure \ref{fig:rcsfrmass_gswlc}) as compared to when spectroscopy-derived SFRs are used (Figure \ref{fig:rcsfr_mass_MPA--JHU}) in the analysis. \\
\\
Having determined that SFR and stellar mass are important for estimating a galaxy's 150\,MHz luminosity, we adopt the stellar-mass dependent form of the relation described in \citetalias{Gurkan2018}:
\begin{equation}\label{L150SFRm_equation}
L_{\textrm{150\,MHz}} = L_c \psi^{\beta} \left( \frac{M_*}{10^{10} {M}_{\odot}} \right)^{\gamma}.
\end{equation}
\noindent where $L_c$ represents the radio luminosity (in W\,Hz$^{-1}$) of a galaxy with stellar mass 10$^{10}$\,$M_\odot$ and SFR = 1\,$M_{\odot}$\,yr$^{-1}$. We use the Markov Chain Monte Carlo (MCMC) algorithm implemented using the \texttt{python} package \texttt{emcee} \citep{emcee}, employing 16 walkers and a chain length of 10,000 samples, to fit the relation. For photometrically-derived SFRs, we obtain the best-fit values, of log $L_{c}$ = 21.838 $\pm$ 0.003, $\beta$ = 1.017 $\pm$ 0.007 and $\gamma$ = 0.311 $\pm$ 0.007. However, for spectroscopy-derived (i.e. MPA--JHU) SFRs, we obtain the best-fit values of log $L_{c}$ = 21.992 $\pm$ 0.004, $\beta$ = 0.644 $\pm$ 0.007 and $\gamma$ = 0.571 $\pm$ 0.008. Taken at face value, the dependence appears very different depending on the origin of the SFR and stellar mass estimates, however \citetalias{Smith20} demonstrated through a series of simulations (see appendix C of \citetalias{Smith20}) that model parameters estimated in this way are biased, and
simulations are required to correct for that bias. In Section \ref{bias}, we follow that work and attempt to correct for this bias. 
\subsection{Bias correction}\label{bias}

\begin{figure*}
    \centering
    \includegraphics[scale=1]{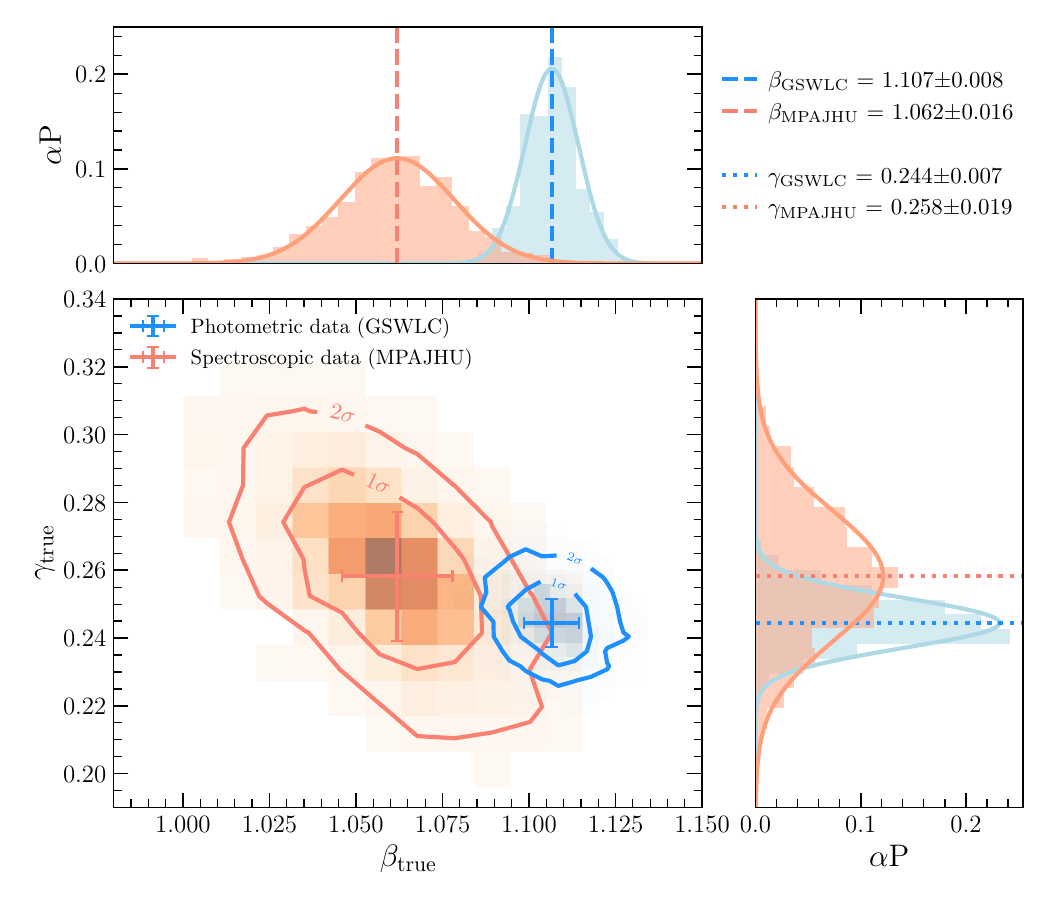}
    \caption{Posterior estimates of bias-corrected $\beta$ and $\gamma$, the indices of the power law terms for SFR and stellar mass in equation \ref{L150SFRm_equation}. Marginal histograms overlaid with best-fit Gaussian curves on the top and right show the 1D PDFs for $\beta$ and $\gamma$ for the MPA--JHU (in orange) and GSWLC (blue) datasets. The dashed lines show the median likelihood values. The 2D distribution of the PDFs is shown on the central plot, with contours representing 68 per cent and 95 per cent confidence intervals. 
    }
    \label{fig:2dplot}
\end{figure*}

On repeating a similar analysis as in \citetalias{Smith20} to correct the best-fit estimates of $\beta$, $\gamma$ and $L_c$ for systematic bias, we find that the degree of bias depends on the noise properties of the values in the dataset even when the uncertainties are included in the fitting. Given that the spectroscopy-derived SFRs and stellar masses from MPA--JHU have larger typical quoted uncertainties ($\bar{\sigma}_{M_*} \approx 0.08\,$dex, and $\bar{\sigma}_{\psi} \approx 0.22$\,dex) compared to those from the photometric GSWLC catalog ($\bar{\sigma}_{M_*} \approx 0.05\,$dex, and 
$\bar{\sigma}_{\psi} \approx 0.08$\,dex), bespoke simulations are required in each particular situation. \\
\\
Following \citetalias{Smith20}, we use simulations described in Appendix \ref{biascorr_expl} to correct for systematic offsets in values of log $L_c$, $\beta$ and $\gamma$ for the stellar mass-dependent model by determining how well we are able to recover known relations of the form given by equation \ref{L150SFRm_equation}. We generate a set of 5000 simulated models of the mass-dependent RC--SFR relation with varying values of $L_c$, $\beta$ and $\gamma$ and then attempt to recover the relation using methods described in Section \ref{rcsfrmass}. The best fit relations between the input and recovered values of $L_c$, $\beta$ and $\gamma$ for the photometric and spectroscopic datasets are given in the Appendix in Section \ref{biascorr_expl} (also shown in Figure \ref{fig:gswlc_truevsrecov}). We then define a quantity $n_{\sigma}$ such that it weights the mock dataset values based on how far away they lie from the observed values of MPA--JHU and GSWLC datasets, i.e. the farther away the points lie from the observed $\beta$ and $\gamma$ best-fit values of the real dataset, the less likely they are to correspond to the true values in the mock datasets: 

\begin{equation}\label{nsigmaequation}
    n_{\sigma} = \sqrt{ \frac{(\beta_m - \beta_r)^2}{\delta_{\beta_m}^2 + \delta_{\beta_r}^2} + \frac{(\gamma_m - \gamma_r)^2}{\delta_{\gamma_m}^2 + \delta_{\gamma_r}^2}}
 \end{equation}

\noindent where $\beta_m$ and $\gamma_m$ represent the observed $\beta$ and $\gamma$ values for the mock datasets, and $\beta_r$ and $\gamma_r$ refer to the observed $\beta$ and $\gamma$ values for the best-fit model using real datasets, i.e. MPA--JHU or GSWLC. The denominators for each term in equation \ref{nsigmaequation} correspond to the sum in quadrature of the uncertainties (the $\delta$ terms) on the recovered values of $\beta$ and $\gamma$ for the mock and real datasets. To obtain plausible estimates of the error budget, we adopt the average scatter across the best-fit lines in Figure \ref{fig:gswlc_truevsrecov}, $\delta_{\beta_m}$,$\delta_{\gamma_m}$ = 0.006,  0.014 (0.003, 0.004) for the spectroscopic (photometric) dataset and $\delta_{\beta_r}$, $\delta_{\gamma_r}$ = 0.007, 0.008 (0.007, 0.007) for the spectroscopic (photometric) dataset, based on the uncertainties derived from fitting equation \ref{L150SFRm_equation} using \texttt{emcee}.\\
\\
We use the derived $n_\sigma$ values to assign weights to all the points in the \textit{true} $\beta$ and $\gamma$ distribution based on their distances between the observed values of $\beta$ and $\gamma$ for the spectroscopic and photometric datasets, estimated from the best fit relations shown in Figure \ref{fig:gswlc_truevsrecov}. These values of $n_\sigma$ are converted to likelihoods from the survival function ($1 -$ CDF) of the $n_\sigma$ which is assumed to be a normal distribution. The results are shown in Figure \ref{fig:2dplot}, where the 2D probability distribution for the $\beta$ and $\gamma$ values is shown in the main panel (overlaid with 1 and 2$\sigma$ contours), plus marginal histograms showing the projected 1D distributions that have been fit with Gaussian curves. \\
\\
Based on the PDFs, we estimate the best-fit values of $\beta$ and $\gamma$ to be 1.107 $\pm$ 0.008 and 0.244 $\pm$ 0.007 for the photometric GSWLC dataset, respectively. For the spectroscopic dataset, we find the corresponding values to be 1.062 $\pm$ 0.016 and 0.258 $\pm$ 0.019. The broader distribution of possible values for the spectroscopic dataset can be attributed to larger uncertainties on the SFRs and stellar masses in the MPA--JHU catalogue relative to those in GSWLC.

\section{Discussion and Conclusions}\label{discussion}

In this work, we studied the dependence of radio luminosity at 150\,MHz on properties of star-forming galaxies (SFGs) and the non-linearity of the low-frequency 150\,MHz RC--SFR relation in local galaxies ($ z \leq 0.3$). Starting from an optically-selected sample of emission-line-classified SFGs, we used a non-parametric random forest regression algorithm to model radio luminosities from the second data release of the LOFAR Low Frequency Two Metre Sky Survey (LoTSS-DR2) given SFRs and stellar masses from the GSWLC-X2 and MPA--JHU DR8 catalogues, and also considering gas-phase metallicity and velocity dispersion measurements from MPA--JHU. Applying this model to a sample of 18,828 sources detected at $\ge 3 \sigma$ in the LoTSS-DR2 catalogue shows that a source's radio luminosity depends on its SFR and to a lesser -- but still highly significant -- extent on its stellar mass. This builds on previous works showing consistent results even when no prior assumption about the form of mass dependence is assumed. These factors remain consistent whether our SFRs and stellar masses were taken from the GSWLC (photometric) or MPA--JHU (spectroscopic) catalogue. We also uncover a possible dependence of a galaxy's 150\,MHz luminosity on the gas-phase metallicity; however, we are unable to make a definitive statement given the well-known degeneracy between stellar mass and [O\slash H]. Having demonstrated the importance of considering stellar mass alongside SFR for this task, we used a suite of simulations to account for measurement bias in estimating the stellar mass dependent RC--SFR relation from \citet{Gurkan2018} using the method of \citet{Smith20}. Our results show that the de-biased best-fitting relation for the photometric dataset (i.e. GSWLC) is given by:

\vspace{-0.05cm}
\begin{align}\label{lsfrm_eq_MPA--JHU}
    \rm{log} (\mathnormal{L}_{150\rm{\,MHz}}/ {\rm W\,Hz^{-1}}) =\ &(1.107 \pm 0.008) \log(\psi/{M}_{\odot}\,\rm{yr}^{-1})\\ \nonumber 
    &+ (0.244\pm 0.007 ) \log(M_*/10^{10}\,{M}_{\odot}) \\
    &+ (21.848 \pm 0.003), \nonumber 
\end{align} 

\noindent and for the spectroscopic (i.e. MPA--JHU) dataset by:
\vspace{-0.05cm}
\begin{align}\label{lsfrm_eq_gswlc}
    \rm{log} (\mathnormal{L}_{150\rm{\,MHz}}/ {\rm W\,Hz^{-1}}) =\ &(1.062 \pm 0.016) \log(\psi/{M}_{\odot}\,\rm{yr}^{-1})\\\nonumber 
    & + (0.258\pm 0.019 ) \log(M_*/10^{10}\,{M}_{\odot}) \\
    &+ (21.910 \pm 0.003) 
\end{align}

\noindent where the values of $L_c$ for the photometric and spectroscopic models are estimated after performing bias correction using equation \ref{lgs} and equation \ref{lmj} respectively. Interestingly, the two sets of measurements are statistically consistent with one another, despite the large differences in the manner in which the SFRs and stellar masses have been estimated in the photometric and spectroscopic datasets (e.g. the SFR estimates alone have different physical time-scales, and the MPA--JHU stellar masses are corrected for nebular emission using the measured spectra, but otherwise rely only on $ugriz$ photometry, as compared with the GSWLC estimates which are derived based on UV to mid-infrared wavelength photometry and energy-balance SED fitting). 
\subsection{Comparison with previous works}
\begin{table*}
    \centering
    \caption{Best-fit values of $L_c$, $\beta$ and $\gamma$ for the parametric model for both photometric (GSWLC) and spectroscopic (MPA--JHU) datasets, compared with those values from previous works.}
   \begin{tabular}{cccc}
        \hline 
         & $L_c$ & $\beta$ & $\gamma$  \\
        \hline 
        \citet{Gurkan2018} & 22.13 $\pm$ 0.01 & 0.77 $\pm$ 0.01 & 0.43 $\pm$ 0.01\\
        \citet{Smith20} & 22.22 $\pm$ 0.02 & 0.90 $\pm$ 0.01& 0.33 $\pm$ 0.04\\
        \citet{das2024} & 22.083 $\pm$ 0.004 & 0.778 $\pm$ 0.004 & 0.334 $\pm$ 0.006\\ 
        This work: Photometric (GSWLC; bias corrected) & 21.848 $\pm$ 0.003 & 1.107 $\pm$ 0.008 & 0.244 $\pm$ 0.007\\
        This work: Spectroscopic (MPA--JHU; bias corrected) & 21.910 $\pm$ 0.003 & 1.062 $\pm$ 0.016 & 0.258 $\pm$ 0.019 \\
         \hline
         \end{tabular}
    \label{tab:bestfitvals}
\end{table*}
The best-fit values of $L_c$, $\beta$ and $\gamma$ for both photometric (GSWLC) and spectroscopic (MPA--JHU) datasets, after bias-correction are summarised in Table \ref{tab:bestfitvals}, along with values from previous works. As shown in the table, we find a steeper dependence of radio luminosity on SFR (i.e. higher $\beta$) and a weaker dependence on stellar mass (i.e. lower value of $\gamma$) compared to the works of \citetalias{Smith20} and \citetalias{das2024}, despite the similar methodology. While one of the differences in this work is the bias correction method, part of the discrepancy may also arise from different sample selections. In contrast to \citetalias{Smith20} and \citetalias{das2024} who use an infrared-selected sample, our sample is drawn from an optically selected spectroscopic catalogue with stringent BPT emission line signal-to-noise cuts as described in Section \ref{selection}. This selection preferentially targets purely SFGs with bright emission lines (and thus high specific SFRs), whilst excluding dusty star-forming systems and optically quiescent galaxies that are present in their works. In addition, those studies make use of the LOFAR Deep Fields, which reach significantly lower flux densities than the wide-area LoTSS survey data used in this work. Consequently, their analyses probe fainter radio sources and lower-SFR galaxies, and the corresponding AGN selection functions are not directly comparable. We also explicitly remove radio-quiet AGN from our sample by using probabilistic classification flags from \citet{drake2024}, mitigating biases where optical SFR indicators are contaminated by the presence of a radiative AGN without corresponding radio excess. The dominant source of radio emission in these quiescent galaxies and radio-quiet AGN remains uncertain \citep{2019PanessaRQAGN}, with several mechanisms proposed, including star-formation (e.g. \citealt{2011Padovani}; \citealt{2015Bonzini}), scaled-down version of powerful jets, outflows (e.g. \citealt{2015Morganti}; \citealt{2019Morabito}), wind-ISM interaction (e.g. \citealt{2022Petley}), and wind-driven shocks \citep{2014Zakamska}. This complexity makes it challenging to isolate a subsample of relatively quiescent galaxies in which radio emission is solely driven by star-formation, given that radio luminosity is also a good tracer of star-formation in the host galaxies of RQ-AGN \citep{2015Bonzini}. Consequently, the presence of additional radio emission not associated with star formation, or underestimated SFRs from dust-obscured star formation can elevate the radio luminosity at fixed SFRs, leading to a flatter observed slope as seen in previous works. At the high SFR regime, our best-fitted relations are consistent with those reported by \citetalias{Gurkan2018}, who similarly used a spectroscopically selected sample. The discrepancy is mainly observed at the low-SFR end, however, this regime is subject to high uncertainty in SFR and stellar mass estimates, given the sensitivity to the choice of star-formation history (e.g. \citealt{2020Lower}, \citealt{2024Haskell}) in SED-fitting and time-scales associated with the SFR tracer \citep{2012KennicuttEvans}, as well as potential contamination from compact RQ-AGN. Taken together, these differences in sample selection and AGN treatment yield a measurement of the RC--SFR relation for purely SFGs, with a correspondingly steeper slope.\\ 
\\
To visualize the best-fit stellar mass-dependent RC--SFR obtained after bias correction, we divide our sample into ten equal-width bins of SFR. For each bin, we calculate the radio luminosity using the median values of SFR and stellar mass in the bin, along with the bias-corrected values of $\beta$ and $\gamma$. These values are represented as blue diamonds for the photometric dataset in Figure \ref{fig:rcsfrmass_gswlc} and orange diamonds for the spectroscopic dataset in Figure \ref{fig:rcsfr_mass_MPA--JHU}. We then draw the best-fit line through these points, indicated by dashed blue and orange lines for the photometric and spectroscopic datasets respectively. For comparison, we also calculate the best-fitted lines based on the mass-dependent relations from \citetalias{Gurkan2018}, \citetalias{Smith20}, and \citetalias{das2024}, and overplot them in the top-left panels of Figures~\ref{fig:rcsfrmass_gswlc} and~\ref{fig:rcsfr_mass_MPA--JHU}. To examine the effect of removing mass dependence, we repeat the procedure using relations with the stellar mass term subtracted, and plot the resulting best-fitted lines in the bottom-left panels of the same figures.\\
\\
\noindent One explanation for the dependence of RC--SFR relation on stellar mass could be the dependence of cosmic ray escape fraction on stellar mass, especially relevant in low-mass galaxies. Massive galaxies have more supernovae and longer cosmic ray escape times, producing higher radio luminosities. \cite{1991Klein} found that dwarf (thus, less massive) galaxies are less efficient in containing cosmic ray electrons, which is consistent with theoretical models (e.g. \citealt{1990ChiWolfendale}; \citealt{1993HelouBicay}; \citealt{NickBeck1997}; \citealt{2008Murphy}; \citealt{2010LackiThompson}; \citealt{2011heesendwarf}). \citetalias{Smith20} state that the stellar mass dependence could also be related to the redshift dependence of the RC--SFR relation but due to the small redshift range of our sample,  it is unlikely that evolution can account for the dependence on stellar mass found in this work, which is similar in magnitude to that observed out to $z = 1$ in \citetalias{Smith20} and \citetalias{das2024}. Similarly, although we consider it unlikely that the stellar mass dependence in our analysis arises from assuming a universal canonical IMF, the potential impact of IMF variations cannot be entirely ruled out. For example, \citet{2011Gunawardhana} report that galaxies with higher SFRs tend to exhibit flatter slopes at the high-mass end of the IMF. Nevertheless, a detailed investigation of the effect of IMF variability lies beyond the scope of this work and given the large statistical size of our sample, such variation is unlikely to appreciably affect our results.\\
\\
Significant care has been taken to remove AGN-dominated galaxies from our sample, but as in previous works (\citetalias{Gurkan2018}, \citetalias{Smith20}, \citetalias{das2024}) it is likely that a number of unidentified AGN remain present. For example, \cite{2024Mezcua} used integral field spectroscopy (IFS) MaNGA\footnote{\url{http://sdss4.org/surveys/manga/}} \citep{2015BundyMaNGA} observations of 2292 dwarf galaxies to show that galaxies which are classified as star-forming using single fibre spectroscopic observations could still have AGN-like ionisation outside the fiber aperture coverage of the fixed small-sized fibre. 17 of the galaxies classified as `SF-AGN'\footnote{Galaxies which are classified as star-forming in the [NII]-H$\alpha$ vs [OIII]/H$\beta$ BPT plot but lie beyond the maximum starburst line in the [SII]/H$\alpha$ vs [OIII]/H$\beta$ plot or the [OI]/H$\alpha$ vs [OIII]/H$\beta$ plots.} by \citet{2024Mezcua} lie in our sample of SFGs, and we also identify a small number of galaxies in our sample that, although classified as SF using the [NII] diagnostic diagram, fall in the `AGN' region of the [SII] and/or [OI] diagnostic diagrams (39 candidates out of 35,099 galaxies). We have verified that our results remain consistent within the quoted uncertainties upon removing the SF-AGN from our sample, but acknowledge that our sample could still contain more unidentified AGN. For example, \cite{2025Arnaudova} investigated the presence of AGN identified using the radio brightness temperature method of \citet{morabito2025} in samples of 150\,MHz sources in the LoTSS deep fields \citep{2023Best}. These sources were identified as SFGs at high confidence using an updated version of the \citet{drake2024} probabilistic source classification methodology. Taken together, it is all but impossible to ensure that residual low-luminosity AGN contamination is not responsible for the mass dependence that we observe in this work. This may change in future as the machinery to combine the different strengths of photometric, spectroscopic and radio morphology information to build a fuller picture of the radio source energy budget is developed.\\
\\
Furthermore, several works have shown that global empirical calibrations like the star-formation rate stellar mass (SFMS) relation \citep{noeske2007,2013Leroy,2016Cano,2021Ellison} and the far-infrared radio correlation (FIRRC; \citealt{2006Hughes}; \citealt{2006Murphy};  \citealt{2010Zhang}; \citealt{2013Tabatabaei}) originate from star-formation activity at local scales. Therefore, it would be useful to also explore the RC--SFR relation on spatially resolved scales using LOFAR radio luminosity maps or from future SKA and LOFAR2.0 maps along with spatially resolved spectra with Integral Field Spectroscopy (IFS) from surveys like SAMI \citep{2012Croom}, MaNGA (see e.g. \citealt{2024jin_rcsfr}) and the William Herschel Telescope Enhanced Area Velocity Explorer \citep[WEAVE;][]{dalton2012,jin2024} large integral field unit (IFU) to better understand the local origins of the empirical calibration. A spatially resolved study could also help mitigate remaining concerns about AGN contamination or shock excitation \citep[e.g.][]{2024Arnaudova}, by enabling the analysis to focus on areas away from the active nucleus. \\
\\
We intend to revisit this topic in future with data from WEAVE, which will spend the upcoming five years performing a suite of surveys including the WEAVE-LOFAR survey \citep{2016Smithweave}. WEAVE-LOFAR will use the WEAVE multi-object spectrograph (MOS) mode to obtain more than a million spectra of sources identified in the LoTSS surveys (both deep and wide), and the RF algorithm presented in this work will be essential for this task. 

\section*{Acknowledgements}
We thank the anonymous reviewer and the editor for their valuable comments and suggestions, which helped improve the clarity and quality of this manuscript. We thank Sugata Kaviraj, Alyssa B. Drake, Akshara Binu, Rafael de Souza, Lingyu Wang and Keerthana Jegatheesan for useful discussions. SS and DJBS acknowledge support from the UK Science and Technology Facilities Council (STFC) via the grant ST/X508408/1. DJBS acknowledges support from the UK STFC via grants ST/V000624/1 and ST/Y001028/1. MIA acknowledges support from the STFC under grant ST/Y000951/1. LKM is grateful for support from a UKRI FLF [MR/Y020405/1]. LRH and DJBS acknowledge support from the STFC in the form of grant ST/Y001028/1. MJH thanks the UK STFC for support [ST/Y001249/1]. SD acknowledges support from the STFC studentship via grant ST/W507490/1. LOFAR \citep{VanHaarlem13} is the Low Frequency Array designed and constructed by ASTRON. It has observing, data processing, and data storage facilities in several countries, which are owned by various parties (each with their own funding sources), and which are collectively operated by the ILT foundation under a joint scientific policy. The ILT resources have benefited from the following recent major funding sources: CNRS-INSU, Observatoire de Paris and Université d'Orléans, France; BMBF, MIWF-NRW, MPG, Germany; Science Foundation Ireland (SFI), Department of Business, Enterprise and Innovation (DBEI), Ireland; NWO, The Netherlands; The Science and Technology Facilities Council, UK; Ministry of Science and Higher Education, Poland; The Istituto Nazionale di Astrofisica (INAF), Italy.  This research made use of the Dutch national e-infrastructure with support of the SURF Cooperative (e-infra 180169) and the LOFAR e-infra group, as well as the University of Hertfordshire high performance computing facility and the LOFAR-UK computing facility located at the University of Hertfordshire and
supported by STFC (ST/V002414/1). The Jülich LOFAR Long Term Archive and the German LOFAR network are both coordinated and operated by the Jülich Supercomputing Centre (JSC), and computing resources on the supercomputer JUWELS at JSC were provided by the Gauss Centre for Supercomputing e.V. (grant CHTB00) through the John von Neumann Institute for Computing (NIC). This research made use of the Italian LOFAR IT computing infrastructure supported and operated by INAF, and by the Physics Department of Turin University (under an agreement with Consorzio Interuniversitario per la Fisica Spaziale) at the C3S Supercomputing Centre, Italy. This work made use of Astropy:\footnote{http://www.astropy.org} a community-developed core Python package and an ecosystem of tools and resources for astronomy \citep{astropy:2013, astropy:2018, 2022astropy}. We acknowledge other Python packages used in this work including \texttt{matplotlib} \citep{Hunter:2007}, \texttt{scikit-learn} \citep{scikit-learn}, \texttt{numpy} \citep{harris2020array}, \texttt{seaborn} \citep{Waskom2021seaborn}, \texttt{scipy} \citep{2020SciPy-NMeth}, \texttt{emcee} \citep{emcee} and \texttt{pandas} \citep{reback2020pandas}.

\section*{Data Availability}
In this work, we made use of radio continuum data from the LoTSS-DR2 catalogue, which can be found here: \url{https://lofar-surveys.org/dr2_release.html}. Spectroscopic and photometric data from the MPA--JHU DR8 catalogue can be found here: \url{https://www.sdss3.org/dr8/spectro/galspec.php} and the photometric SED-fit data from GSWLC-X2 can be found here: \url{https://salims.pages.iu.edu/gswlc/}. The exact sample used in this work after selection cuts, will be made available upon request to the primary author.


\bibliographystyle{mnras}
\bibliography{refs}



\appendix
\section{Dependence on Morphological features}\label{morphappendix}
\begin{figure*}
    \centering
    \includegraphics[scale=0.5]{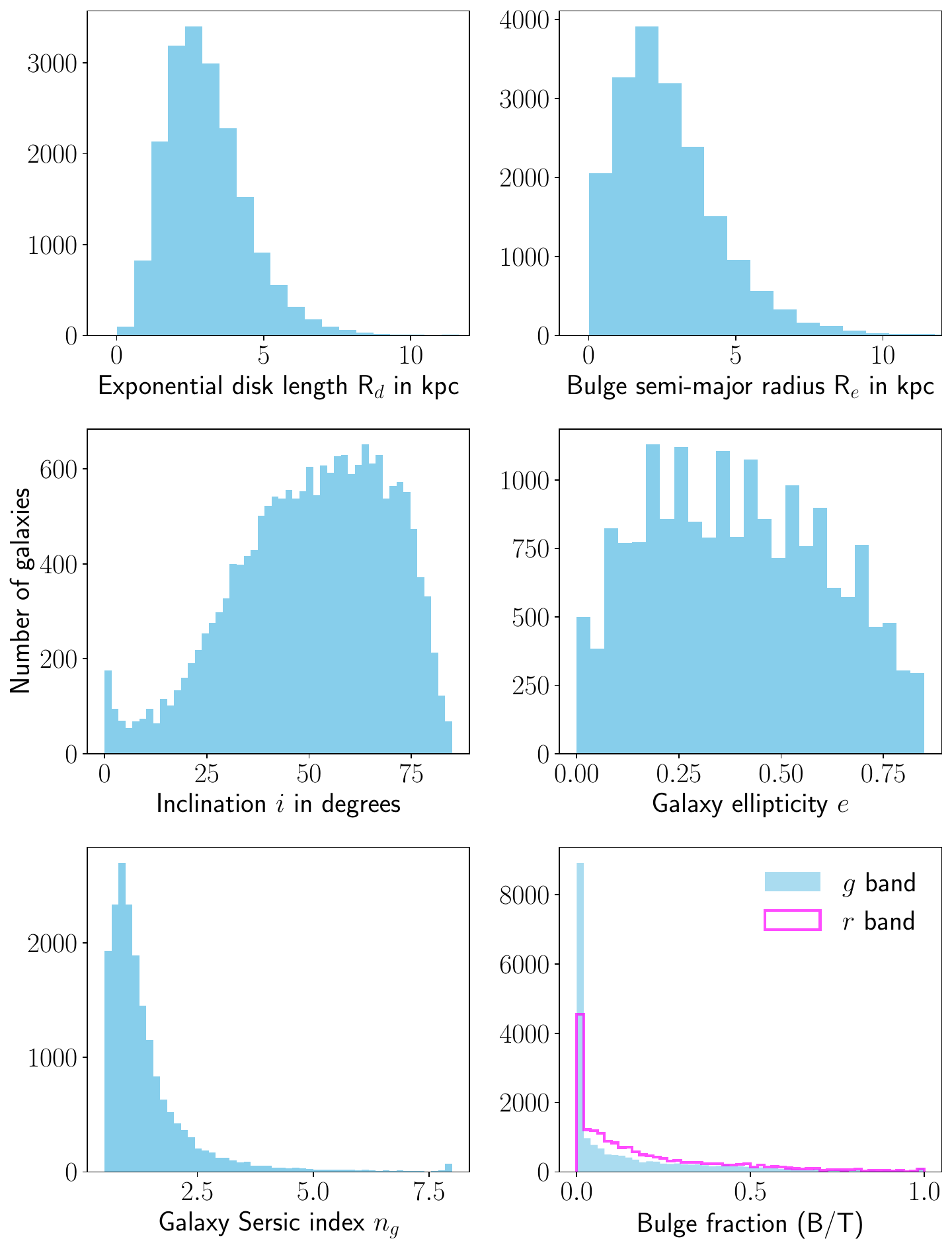}
    \caption{Distribution of the morphological features used in this work. The top row contains exponential disc length and bulge semi major radius in kpc. The middle row shows inclination and galaxy ellipticity. The bottom row shows the distribution of Sersic index and bulge fraction for our sample. The bulge fraction shown in magenta is measured using the SDSS $r-$band images.}
    \label{fig:morphhist}
\end{figure*}

\begin{figure*}
    \centering
    \includegraphics[scale=0.7]{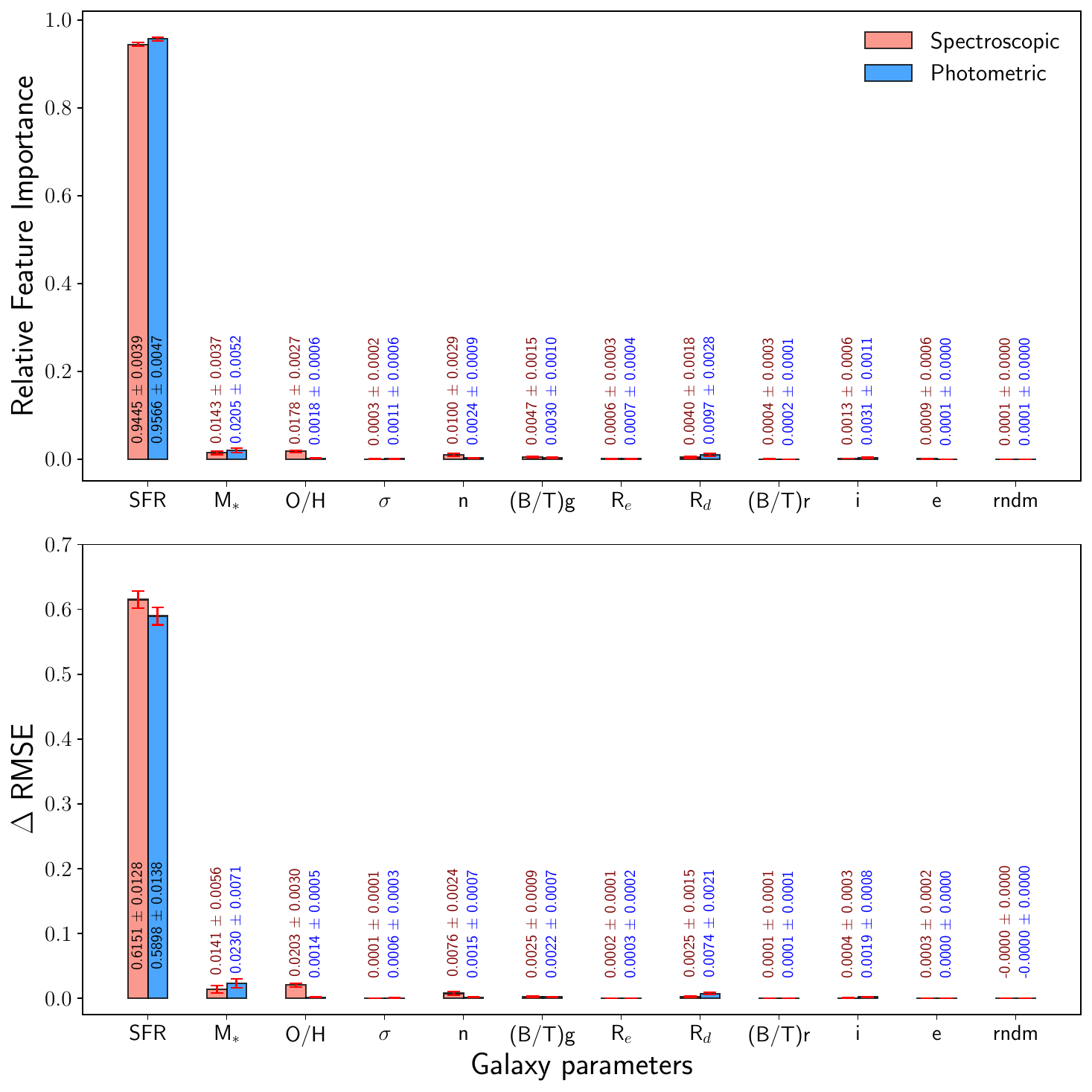}
    \caption{\textit{Upper panel:} Random forest feature importance of SFR, stellar mass, gas-phase metallicity, velocity dispersion and morphological features used in the regression model to predict radio luminosity. \textit{Lower panel:} Same features as the upper panel, but the bars indicate the permuted feature importance, i.e. the decrease in RMSE of the model prediction when the feature is randomly permuted. In both panels, the orange colour indicates SFR and stellar mass taken from the MPA--JHU catalogue and blue indicates SFR and stellar mass taken from the GSWLC, estimated using SED-fitting. }
    \label{figd:morph_imps}
\end{figure*}

Among the reported morphological parameters in \cite{2011Simard} for SDSS-observed galaxies, we consider the following features estimated using pure Sersic decompositions and $n=4$ bulge+disc decomposition (as given in Tables 1 and 3 of \citealt{2011Simard}):
\begin{enumerate}
    \item galaxy ellipticity (\textit{${e}$})
    \item inclination (${\textit{i}}$; \textit{${i=0}$} for face-on disc)
    \item bulge-to-total ratio $B/T$ in the SDSS ${\textit{g}}$ and ${\textit{r}}$ bands 
    \item exponential disc scale length ($R_d$, in kpc)
    \item bulge semi-major effective radius ($R_e$, in kpc)
    \item Sersic index \textit{$n_g$}
\end{enumerate}
and include these in the RF regression analysis along with SFR, stellar mass, gas-phase metallicity and velocity dispersion, for the spectroscopic as well as photometric models. The distribution of these morphological features for our catalogue sample is shown in Fig. \ref{fig:morphhist}. On running the RF model, we find that SFR is still the most important feature in predicting radio luminosity (see Fig. \ref{figd:morph_imps}), for the photometric as well as spectroscopic models. Furthermore, the overall prediction accuracy of the model remains similar to the models in Section \ref{ml} (MSE $\sim$ 0.11 and accuracy $\sim$ 81\ per cent), upon including the morphological parameters. We observe non-negligible feature importance values for some features, e.g. Sersic index in the spectroscopic model and disc length $R_d$ in the photometric model. However, it is difficult to ascertain if they have real physical significance or if this is due to their correlation with stellar mass, solely based on RF regression analysis, given the limitations of RF regression analysis in handling collinear variables (i.e. when two or more variables are highly correlated).

\section{Bias correction using simulations}\label{biascorr_expl}

\begin{figure}
    \centering
    \includegraphics[width=0.45\textwidth]{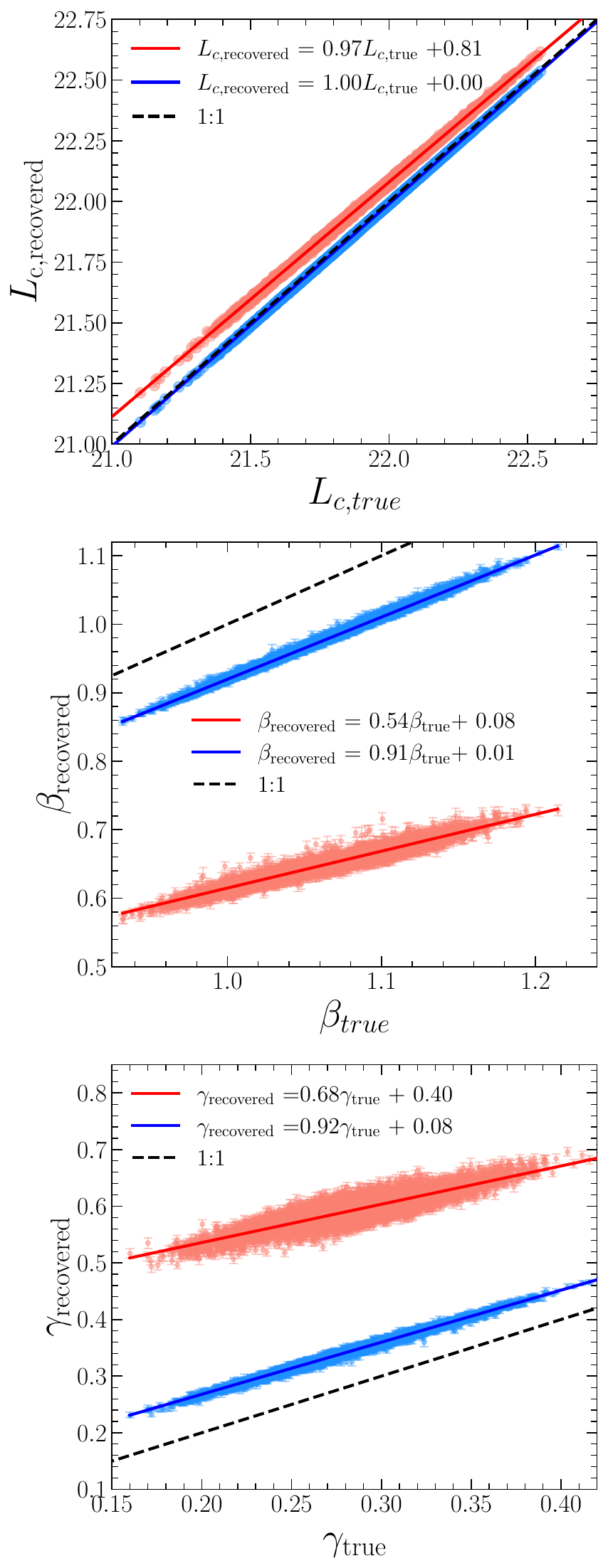}
    \captionof{figure}{Recovered vs true values of $L_c$, $\beta$ and $\gamma$ for the stellar mass dependent form of the RC--SFR for the mock datasets made using photometric data from the GSWLC catalogue (shown in blue) and spectroscopic data from the MPA--JHU catalogue (shown in orange). The best-fitting lines are shown by the solid blue (for the photometric dataset) and red (for the spectroscopic dataset) lines, and the black dashed line is the line of equality.} 
    \label{fig:gswlc_truevsrecov}
\end{figure}

\citetalias{Smith20} showed that the best-fitting estimates of $\beta$ and $\gamma$ in equation \ref{L150SFRm_equation} are likely to be offset by a residual bias which we found to be dependent on the noise properties of the parameters used (here, SFRs and stellar mass) in this work. To correct for this bias, we generated realistic mock data for each set of parameters, i.e. radio flux densities, SFRs and stellar mass values for the photometric as well as spectroscopic datasets through the procedure described in detail in \citetalias{Smith20}. We then fit the mock galaxy data in the same manner as described in Section \ref{rcsfrmass} and plot the recovered versus true values of $L_c$, $\beta$ and $\gamma$ as shown in Fig. \ref{fig:gswlc_truevsrecov} for the photometric and spectroscopic datasets in blue and orange colours respectively. The best-fitting relations between the recovered and true values of $L_c$, $\beta$ and $\gamma$ for the photometric dataset are given by:
    
\begin{equation}\label{lgs}
L_{c,\rm{rec}} = 1.00 L_{c,\rm{true}} + 0.00 
\end{equation}
\begin{equation}
\beta_{\rm{rec}} = 0.91 \beta_{\rm{true}} + 0.01
\end{equation}
\begin{equation}
\gamma_{\rm{rec}} = 0.92 \gamma_{\rm{true}} + 0.08   
\end{equation}

\noindent and for the spectroscopic dataset:

\begin{equation}\label{lmj}
L_{c,\rm{rec}} = 0.97 L_{c,\rm{true}} + 0.81
\end{equation}
\begin{equation}
\beta_{\rm{rec}} = 0.54 \beta_{\rm{true}} + 0.08
\end{equation}
\begin{equation}
\gamma_{\rm{rec}} = 0.68 \gamma_{\rm{true}} + 0.40 
\end{equation}

\bsp	
\label{lastpage}
\end{document}